\documentclass[useAMS,usenatbib,letterpaper]{mn2e}

\usepackage{amsmath}
\usepackage{graphicx}
\usepackage{amssymb}

\title[Wavelets on the sphere]
{Exact reconstruction with directional wavelets on the sphere}

\author[Wiaux et al.]
{Y. Wiaux$^{1}$\thanks{E-mail: yves.wiaux@epfl.ch}, J. D. McEwen$^{2}$\thanks{E-mail: mcewen@mrao.cam.ac.uk}, P. Vandergheynst$^{1}$\thanks{E-mail: pierre.vandergheynst@epfl.ch}, O. Blanc$^{1}$\\
$^{1}$Institute of Electrical Engineering, Ecole Polytechnique F\'ed\'erale de Lausanne (EPFL), CH-1015 Lausanne, Switzerland\\
$^{2}$Astrophysics Group, Cavendish Laboratory, University of Cambridge, Cambridge CB3 0HE, United Kingdom\\}

\begin{document}

\date{\today}

\pagerange{\pageref{firstpage}--\pageref{lastpage}} \pubyear{2007}

\maketitle

\label{firstpage}

\begin{abstract}
A new formalism is derived for the analysis and exact reconstruction
of band-limited signals on the sphere with directional wavelets. It
represents an evolution of the wavelet formalism developed by \citet{antoine99}
and \citet{wiaux05}. The translations of the wavelets at any point
on the sphere and their proper rotations are still defined through
the continuous three-dimensional rotations. The dilations of the wavelets
are directly defined in harmonic space through a new kernel dilation,
which is a modification of an existing harmonic dilation. A family
of factorized steerable functions with compact harmonic support which
are suitable for this kernel dilation is firstly identified. A scale
discretized wavelet formalism is then derived, relying on this dilation.
The discrete nature of the analysis scales allows the exact reconstruction
of band-limited signals. A corresponding exact multi-resolution algorithm
is finally described and an implementation is tested. The formalism
is of interest notably for the denoising or the deconvolution of signals
on the sphere with a sparse expansion in wavelets. In astrophysics,
it finds a particular application for the identification of localized
directional features in the cosmic microwave background (CMB) data,
such as the imprint of topological defects, in particular cosmic strings,
and for their reconstruction after separation from the other signal
components.
\end{abstract}


\section{Introduction}

\label{sec:Introduction}Very generically, the scale-space analysis
of a signal with wavelets on a given manifold defines wavelet coefficients
which characterize the signal around each point of the manifold and
at various scales \citep{mallat98,antoine04,antoine07}. Wavelet techniques
find numerous applications in astrophysics \citep{starck06a}. It
commonly concerns the analysis of data distributed on the real line
of time, or images on the plane. But other experiments also acquire
data in all directions of the sky. This is notably the case of observations
of the cosmic microwave background (CMB) radiation, such as the current
Wilkinson Microwave Anisotropy Probe (WMAP) satellite experiment,
or the forthcoming Planck Surveyor satellite experiment. Sky surveys
such as the NRAO Very Large Array Sky Survey (NVSS) also map data
on a large fraction the celestial sphere. The scale-space analysis
of such data sets requires wavelet techniques on the sphere. Various
wavelet formalisms have been proposed to date \citep{holschneider96,freeden96,freeden98,antoine98,antoine99,narcowich05,mcewen06}.
The formalism originated by \citet{antoine99} in a group-theoretic
context triggered various developments \citep{antoine02,demanet03,bogdanova05},
and was reconsidered in a more practical context by \citet{wiaux05}.
This approach notably found a recent and very interesting application
in the analysis of the CMB data, as reviewed by \citet{mcewen07b}.

In particular, the denoising or the deconvolution of data represents
a large field of application of wavelet techniques. Experimental data
sets are indeed always affected by various noise sources, notably
related to the instrumentation. The data can also be blurred by experimental
beams associated with the instrumentation. Signals detected can also
originate from different physical sources, which need to be separated.
In that component separation perspective, each component can in turn
be technically understood as a signal, while the other components
are seen as noise. As an example, observed CMB data represent a superposition
of the CMB signal itself with instrumental noise and foreground emissions,
blurred by the experimental beam at each detection frequency. The
denoising and the deconvolution of the signal, and the separation
of its astrophysical components is of major interest for astrophysics
and cosmology.

Signals with features defined at specific positions and scales typically
have a sparse expansion in terms of wavelets. For such signals, denoising
and deconvolution algorithms are generically much more efficient when
applied to the wavelet coefficients \citep{mallat98,daubechies04}.
However this requires a scheme allowing the exact reconstruction of
the signals analyzed from their wavelet coefficients. Moreover, localized
characteristics can be elongated, in which case directional wavelets
are essential. The identification and the reconstruction of localized
directional features in CMB data represents a very interesting application
of such a framework. It typically concerns the imprint of topological
defects such as textures or cosmic strings \citep{kaiser84,turok90,vilenkin94,hindmarsh95}.
This application can be recast in a component separation approach
where all continuous and typically Gaussian emissions are seen as
noise, in contrast with localized directional features. Let us also
emphasize that such a framework can have many applications well beyond
astrophysics, from geophysics to biomedical imaging, or computer vision.

At present, the simultaneous combination of the properties of exact
reconstruction and directionality is lacking in the existing wavelet
formalisms on the sphere. It has only been considered for the wavelet
analysis of signals on the plane \citep{simoncelli92,vandergheynst02}.
The primary aim of the present work resides in the development of
a new scale discretized wavelet formalism for the analysis and exact
reconstruction of band-limited signals on the sphere with directional
wavelets. As a by-product, a new continuous wavelet formalism is also
obtained, which allows the analysis of signals with a new family of
wavelets relative to existing formalisms. But the continuous range
of scales required for the analysis prevents exact reconstruction
in practice, for which the scale discretized wavelet formalism proposed
is essential.

The remainder of this paper is organized as follows. In Section \ref{sec:Continuous-wavelets},
we present an existing scheme for the definition of a continuous wavelet
formalism on the sphere from a generic dilation operation. We consider
directional and axisymmetric wavelets and discuss the cases of the
stereographic and harmonic dilations. In Section \ref{sec:Kernel-dilation},
we propose a new kernel dilation. A corresponding family of factorized
steerable functions with compact harmonic support is identified. We
show that localization and directionality properties of such functions
can be controlled through kernel dilation. In Section \ref{sec:Scale-discretized-wavelets},
we derive a new continuous wavelet formalism from the kernel dilation
with continuous scales. We then derive a new scale discretized wavelet
formalism that allows the exact reconstruction of band-limited signals
in practice. We design explicitly an example wavelet. We finally recast
the scale discretized wavelet formalism in an invertible filter bank
approach. In Section \ref{sec:Multiresolution-algorithm}, we describe
an exact algorithm accounting for the multi-resolution properties
of the formalism. The memory and computation time requirements are
discussed and an implementation is tested. In Section \ref{sec:Astrophysical-application},
we discuss the application of the formalism to the detection of cosmic
strings through the denoising of full-sky CMB data. We finally conclude
in Section \ref{sec:Conclusion}.

\section{Wavelets from a generic dilation}

\label{sec:Continuous-wavelets}In this section we discuss an existing
scheme for the definition of a continuous wavelet formalism on the
sphere from a generic dilation operation. We consider directional
and axisymmetric wavelets explicitly. We finally discuss in detail
the stereographic and harmonic dilations.

\subsection{Directional wavelets}

\label{sub:Directional-wavelets}In the continuous framework developed
by \citet{antoine99} and \citet{wiaux05}, the wavelet analysis of
a signal on the sphere\emph{, i.e.} the unit sphere $\textnormal{S}^{2}$,
defines wavelet coefficients through the correlation of the signal
with dilated versions of a local analysis function. Theoretically,
the signal can be recovered explicitly from its wavelet coefficients
provided that the local analysis functions satisfies some admissibility
condition, raising it to the rank of a wavelet.

The real and harmonic structures of $\textnormal{S}^{2}$ are summarized
concisely as follows. We consider a three-dimensional Cartesian coordinate
system $(o,o\hat{x},o\hat{y},o\hat{z})$ centered on the sphere, and
where the direction $o\hat{z}$ identifies the North pole. Any point
$\omega$ on the sphere is identified by its corresponding spherical
coordinates $(\theta,\varphi)$, where $\theta\in[0,\pi]$ stands
for the co-latitude, or polar angle, and $\varphi\in[0,2\pi)$ for
the longitude, or azimuthal angle. Let the continuous signal $F(\omega)$
and the local analysis function $\Psi(\omega)$ be square-integrable
functions on the sphere: $F,\Psi\in\textnormal{L}^{2}(\textnormal{S}^{2},\textnormal{d}\Omega)$,
with the invariant measure $\textnormal{d}\Omega=\textnormal{d}\cos\theta\textnormal{d}\varphi$.
The spherical harmonics form an orthonormal basis for the decomposition
of square-integrable functions. They are explicitly given in a factorized
form in terms of the associated Legendre polynomials $P_{l}^{m}(\cos\theta)$
and the complex exponentials $e^{im\varphi}$ as \begin{equation}
Y_{lm}\left(\theta,\varphi\right)=\left[\frac{2l+1}{4\pi}\frac{\left(l-m\right)!}{\left(l+m\right)!}\right]^{1/2}P_{l}^{m}\left(\cos\theta\right)e^{im\varphi},\label{eq:cw1}\end{equation}
with $l\in\mathbb{N}$, $m\in\mathbb{Z}$, and $|m|\leq l$ \citep{abramowitz65,varshalovich89}.
The index $l$ represents an overall frequency on the sphere. The
absolute value $\vert m\vert$ represents the frequency associated
with the azimuthal variable $\varphi$. Any function $G\in\textnormal{L}^{2}(\textnormal{S}^{2},\textnormal{d}\Omega)$
is thus uniquely given as a linear combination of scalar spherical
harmonics: $G(\omega)=\sum_{l\in\mathbb{N}}\sum_{|m|\leq l}\widehat{G}_{lm}Y_{lm}(\omega)$.
This combination defines the inverse spherical harmonic transform
on $\textnormal{S}^{2}$. The corresponding spherical harmonic coefficients
are given by the projection $\widehat{G}_{lm}=\langle Y_{lm}\vert G\rangle$,
with $\vert m\vert\leq l$, where the bracket $\langle F_{2}\vert F_{1}\rangle=\int_{\textnormal{S}^{2}}\textnormal{d}\Omega\, F_{2}^{*}(\omega)F_{1}(\omega)$
generically denotes the scalar product for $F_{1},F_{2}\in\textnormal{L}^{2}(\textnormal{S}^{2},\textnormal{d}\Omega)$.
This projection defines the direct spherical harmonic transform on
$\textnormal{S}^{2}$.

Continuous affine transformations such as translations, rotations,
and dilations are applied to the analysis function. The continuous
translations by $\omega_{0}=(\theta_{0},\varphi_{0})\in\textnormal{S}^{2}$
and rotations by $\chi\in[0,2\pi)$ are defined by the three Euler
angles defining an element $\rho=(\varphi_{0},\theta_{0},\chi)$ of
the group of rotations in three dimensions $SO(3)$. The operator
$R(\omega_{0})$ in $\textnormal{L}^{2}(\textnormal{S}^{2},\textnormal{d}\Omega)$
for the translation of amplitude $\omega_{0}=(\theta_{0},\varphi_{0})$
of a function $G$ reads as \begin{equation}
G_{\omega_{0}}\left(\omega\right)=\left[R\left(\omega_{0}\right)G\right]\left(\omega\right)=G\left(R_{\omega_{0}}^{-1}\omega\right),\label{eq:cw2}\end{equation}
where $R_{\omega_{0}}(\theta,\varphi)=[R_{\varphi_{0}}^{\hat{z}}R_{\theta_{0}}^{\hat{y}}](\theta,\varphi)$
is defined by the three-dimensional rotation matrices $R_{\theta_{0}}^{\hat{y}}$
and $R_{\varphi_{0}}^{\hat{z}}$, acting on the Cartesian coordinates
$(x,y,z)$ associated with $\omega=(\theta,\varphi)$. The rotation
operator $R^{\hat{z}}(\chi)$ in $\textnormal{L}^{2}(\textnormal{S}^{2},\textnormal{d}\Omega)$
for the rotation of the function $G$ around itself, by an angle $\chi\in[0,2\pi)$,
is given as \begin{equation}
G_{\chi}\left(\omega\right)=\left[R^{\hat{z}}\left(\chi\right)G\right]\left(\omega\right)=G\left({R_{\chi}^{\hat{z}}}^{-1}\omega\right),\label{eq:cw3}\end{equation}
where $R_{\chi}^{\hat{z}}(\theta,\varphi)=(\theta,\varphi+\chi)$
also follows from the action of the three-dimensional rotation matrix
$R_{\chi}^{\hat{z}}$ on the Cartesian coordinates $(x,y,z)$ associated
with $\omega=(\theta,\varphi)$. The operator incorporating both the
translations and rotations simply reads as $R(\rho)=R(\omega_{0})R^{\hat{z}}(\chi)$
and $G_{\rho}(\omega)=[R(\rho)G](\omega)=G(R_{\rho}^{-1}\omega)$,
with $R_{\rho}=R_{\omega_{0}}R_{\chi}^{\hat{z}}$. The continuous
dilations affect by definition the continuous scale of the function.
The notion of scale may \emph{a priori} be defined both in real or
in harmonic space on $\textnormal{S}^{2}$. In the remainder of the
present subsection we simply denote the dilated function as $G_{a}(\omega)$,
where $a\in\mathbb{R}_{+}^{*}$ stands for a continuous dilation factor.
We explicitly discuss two possible definitions of dilations in Subsections
\ref{sub:Stereographic-dilation} and \ref{sub:Harmonic-dilation}.

The analysis of the signal $F$ with an analysis function $\Psi$
defines wavelet coefficients through the directional correlation of
$F$ with the dilated functions $\Psi_{a}$, \emph{i.e.} the scalar
products\begin{equation}
W_{\Psi}^{F}\left(\rho,a\right)=\langle\Psi_{\rho,a}|F\rangle.\label{eq:cw4}\end{equation}
At each scale $a$, the wavelet coefficients $W_{\Psi}^{F}(\rho,a)$
therefore identify a square-integrable function on the rotation group
in three dimensions $\textnormal{SO(3)}$. They characterize the signal
around each point $\omega_{0}$, and in each orientation $\chi$.
This defines the scale-space nature of the wavelet decomposition on
the sphere.

The real and harmonic structures of the rotation group in three dimensions
$\textnormal{SO(3)}$ are summarized concisely as follows. As discussed,
any rotation $\rho$ on $\textnormal{SO(3)}$ is given in terms of
the three Euler angles $\rho=(\varphi,\theta,\chi)$, with $\theta\in[0,\pi]$,
and $\varphi,\chi\in[0,2\pi)$. Let $H(\rho)$ be a square-integrable
function on $\textnormal{SO(3)}$: $H\in\textnormal{L}^{2}(\textnormal{SO(3)},\textnormal{d}\rho)$,
with the invariant measure $\textnormal{d}\rho=\textnormal{d}\varphi\textnormal{d}\cos\theta\textnormal{d}\chi$.
The Wigner $D$-functions are the matrix elements of the irreducible
unitary representations of weight $l$ of the group in $\textnormal{L}^{2}(\textnormal{SO(3)},\textnormal{d}\rho)$.
By the Peter-Weyl theorem on compact groups, the matrix elements $D_{mn}^{l*}$
also form an orthogonal basis in $\textnormal{L}^{2}(\textnormal{SO(3)},\textnormal{d}\rho)$.
They are explicitly given in a factorized form in terms of the real
Wigner $d$-functions $d_{mn}^{l}(\theta)$ and the complex exponentials,
$e^{-im\varphi}$ and $e^{-in\chi}$, as \begin{equation}
D_{mn}^{l}\left(\varphi,\theta,\chi\right)=e^{-im\varphi}d_{mn}^{l}\left(\theta\right)e^{-in\chi},\label{eq:cw5}\end{equation}
 with $l\in\mathbb{N}$, $m,n\in\mathbb{Z}$, and $|m|,|n|\leq l$
\citep{varshalovich89,brink93}. Again, $l$ represents an overall
frequency on $\textnormal{SO(3)}$, and $\vert m\vert$ and $\vert n\vert$
the frequencies associated with the variables $\varphi$ and $\chi$,
respectively. Any function $H\in\textnormal{L}^{2}(\textnormal{SO(3)},\textnormal{d}\rho)$
is thus uniquely given as a linear combination of Wigner $D$-functions:
$H(\rho)=\sum_{l\in\mathbb{N}}(2l+1)/8\pi^{2}\sum_{|m|,|n|\leq l}\widehat{H}_{mn}^{l}D_{mn}^{l*}(\rho)$.
This combination defines the inverse Wigner $D$-function transform
on $\textnormal{SO(3)}$. The corresponding Wigner $D$-function coefficients
are given by the projection $\widehat{H}_{mn}^{l}=\int_{\textnormal{SO(3)}}\textnormal{d}\rho\, D_{mn}^{l}(\rho)H(\rho)$.
This projection defines the direct Wigner $D$-function transform
on $\textnormal{SO(3)}$.

At each scale, the direct Wigner $D$-function transform of the wavelet
coefficients is given as the pointwise product of the spherical harmonic
coefficients of the signal and the wavelet:\begin{equation}
\widehat{\left(W_{\Psi}^{F}\right)}_{mn}^{l}\left(a\right)=\frac{8\pi^{2}}{2l+1}\widehat{\left(\Psi_{a}\right)}_{ln}^{*}\widehat{F}_{lm}.\label{eq:cw6}\end{equation}
Indeed, the orthonormality of scalar spherical harmonics implies the
Plancherel relation $\langle F_{2}\vert F_{1}\rangle=\sum_{l\in\mathbb{N}}\sum_{|m|\leq l}\widehat{(F_{2})}_{lm}^{*}\widehat{(F_{1})}_{lm}$
for $F_{1},F_{2}\in\textnormal{L}^{2}(\textnormal{S}^{2},\textnormal{d}\Omega)$,
and the action of the operator $R(\rho)$ on $G\in\textnormal{L}^{2}(\textnormal{S}^{2},\textnormal{d}\Omega)$
reads in terms of its spherical harmonic coefficients as $\widehat{(G_{\rho})}_{lm}=\sum_{|n|\leq l}D_{mn}^{l}(\rho)\widehat{G}_{ln}.$

The reconstruction of a signal $F$ from its wavelet coefficients
with an analysis function $\Psi$ is given as \begin{equation}
F\left(\omega\right)=\int_{\mathbb{R}_{+}^{*}}\textnormal{d}\mu\left(a\right)\int_{\textnormal{SO(3)}}\textnormal{d}\rho\, W_{\Psi}^{F}\left(\rho,a\right)\left[R\left(\rho\right)L_{\Psi}\Psi_{a}\right]\left(\omega\right).\label{eq:cw7}\end{equation}
In this relation, the scale integration measure $\textnormal{d}\mu(a)$
is part of the definition of the dilation operation itself (see Subsections
\ref{sub:Stereographic-dilation} and \ref{sub:Harmonic-dilation}).
The operator $L_{\Psi}$ in $\textnormal{L}^{2}(\textnormal{S}^{2},\textnormal{d}\Omega)$
is defined by its action on the spherical harmonic coefficients of
a function $G$: $\widehat{L_{\Psi}G}_{lm}=\widehat{G}_{lm}/C_{\Psi}^{l}$.
The reconstruction formula holds if and only if the analysis function
satisfy the following admissibility condition for all $l\in\mathbb{N}$:
\begin{equation}
0<C_{\Psi}^{l}=\frac{8\pi^{2}}{2l+1}\sum_{|m|\leq l}\int_{\mathbb{R}_{+}^{*}}\textnormal{d}\mu\left(a\right)\,|\widehat{\left(\Psi_{a}\right)}_{lm}|^{2}<\infty.\label{eq:cw8}\end{equation}
In this case the analysis function $\Psi$ is by definition raised
to the rank of a wavelet. From relation (\ref{eq:cw8}), the admissibility
condition intuitively requires that the whole wavelet family \{$\Psi_{a}(\omega)$\},
for $a\in\mathbb{R}_{+}^{*}$, covers each frequency index $l$ with
a finite and non-zero amplitude, hence preserving the signal information
at each frequency. Notice that a direct connection exists between
the generic relations (\ref{eq:cw7}) and (\ref{eq:cw8}) for the
signal reconstruction, and the theory of frames on the sphere \citep{bogdanova05}.

We generally consider band-limited signals. Any function $G\in\textnormal{L}^{2}(\textnormal{S}^{2},\textnormal{d}\Omega)$
is said to be band-limited with band limit $B$, for any $B\in\mathbb{N}^{0}$,
if $\widehat{G}_{lm}=0$ for all $l,m$ with $l\geq B$. Any function
$H\in\textnormal{L}^{2}(\textnormal{SO(3)},\textnormal{d}\rho)$ is
said to be band-limited with band limit $B$, for any $B\in\mathbb{N}^{0}$,
if $\widehat{H}_{mn}^{l}=0$ for all $l,m,n$ with $l\geq B$. From
relation (\ref{eq:cw6}), if the signal $F$ or the wavelet $\Psi$
are band-limited on $\textnormal{S}^{2}$, then the wavelet coefficients
$W_{\Psi}^{F}$ are automatically band-limited on $\textnormal{SO(3)}$,
with the same band limit $B$.

Let us already notice that the reconstruction is ensured theoretically
from relation (\ref{eq:cw7}), through the integration on the continuous
parameter $\rho\in\textnormal{SO(3)}$ for translations and rotations
of the wavelet, and on the continuous dilation factor $a\in\mathbb{R}_{+}^{*}$.
But in practice, the reconstruction would require the definition of
exact quadrature rules for the numerical integrations. Exact quadrature
rules for integration of band-limited signals on $\textnormal{S}^{2}$
exist on equi-angular \citep{driscoll94} and Gauss-Legendre \citep{doroshkevich05a,doroshkevich05b}
pixelizations of $(\theta_{0},\varphi_{0})$. HEALPix pixelizations
\citep{gorski05}%
\footnote{http://healpix.jpl.nasa.gov/%
} of $(\theta_{0},\varphi_{0})$ on $\textnormal{S}^{2}$ provide approximate
quadrature rules which can also be made very precise thanks to an
iteration process. Pixelizations may for instance be defined on $SO(3)$
by combining pixelizations on $\textnormal{S}^{2}$ with an equi-angular
sampling of $\chi$. Corresponding quadrature rules can be made exact
on the pixelizations based on equi-angular and Gauss-Legendre pixelizations
on $\textnormal{S}^{2}$, while those based on HEALPix pixelizations
are approximate. This extension basically relies on the separation
of the integration variables \citep{maslen97a,maslen97b,kostelec03}
from relation (\ref{eq:cw5}).

However, exact quadrature rules do not exist for the integration over
scales $a\in\mathbb{R}_{+}^{*}$. In practice, this prevents an exact
reconstruction of the signal analyzed. A scheme allowing an exact
reconstruction requires a discretization of the dilation factor. A
scale discretized wavelet formalism is proposed in Section \ref{sec:Scale-discretized-wavelets},
thanks to a specific choice of dilation, and through an integration
of the dilation factor by slices in relation (\ref{eq:cw7}).

\subsection{Axisymmetric wavelets}

\label{sub:Axisymmetric-case}Any general function $G\in\textnormal{L}^{2}(\textnormal{S}^{2},\textnormal{d}\Omega)$
explicitly dependent on the azimuthal angle $\varphi$, is said to
be directional: $G=G(\theta,\varphi)$. By opposition, any function
$A\in\textnormal{L}^{2}(\textnormal{S}^{2},\textnormal{d}\Omega)$
independent of the azimuthal angle $\varphi$ is said to be zonal,
or axisymmetric: $A=A(\theta)$. It only exhibits non-zero spherical
harmonic coefficients for $m=0$: $\widehat{A}_{lm}=\widehat{A}_{l0}\delta_{m0}$.

In this particular case, the directional correlation of a signal $F$
with $A$ reduces to a standard correlation obviously independent
of the rotation angle $\chi$ \citep{wiaux06}. The analysis of $F$
with an axisymmetric analysis function $A$ defines wavelet coefficients
through the standard correlation of $F$ with the dilated functions
$A_{a}$, \emph{i.e.} the scalar products\begin{equation}
W_{A}^{F}\left(\omega_{0},a\right)=\langle A_{\omega_{0},a}|F\rangle.\label{eq:cw9}\end{equation}
At each scale $a$, the wavelet coefficients identify a square-integrable
function on $\textnormal{S}^{2}$ rather than on $\textnormal{SO(3)}$.
The spherical harmonic transform of the wavelet coefficients is still
given as the pointwise product of the spherical harmonic coefficients
of the signal and the wavelet:\begin{equation}
\widehat{\left(W_{A}^{F}\right)}_{lm}\left(a\right)=\sqrt{\frac{4\pi}{2l+1}}\widehat{\left(A_{a}\right)}_{l0}^{*}\widehat{F}_{lm}.\label{eq:cw10}\end{equation}
This relation simply follows from relation (\ref{eq:cw6}) and the
equality $D_{m0}^{l}(\omega,0)=[4\pi/(2l+1)]^{1/2}Y_{lm}^{*}(\omega)$.

The reconstruction of $F$ from its wavelet coefficients reads as:
\begin{equation}
F\left(\omega\right)=\int_{\mathbb{R}_{+}^{*}}\textnormal{d}\mu\left(a\right)\int_{\textnormal{S}^{2}}\textnormal{d}\omega_{0}\, W_{A}^{F}\left(\omega_{0},a\right)\left[R\left(\omega_{0}\right)L_{A}A_{a}\right]\left(\omega\right),\label{eq:cw11}\end{equation}
for any scale integration measure $\textnormal{d}\mu(a)$, and with
the operator $L_{A}$ in $\textnormal{L}^{2}(\textnormal{S}^{2},\textnormal{d}\Omega)$
defined by: $\widehat{L_{A}G}_{l0}=\widehat{G}_{l0}/C_{A}^{l}$. The
reconstruction formula holds if and only if the analysis function
satisfies the following admissibility condition for all $l\in\mathbb{N}$:
\begin{equation}
0<C_{A}^{l}=\frac{4\pi}{2l+1}\int_{\mathbb{R}_{+}^{*}}\textnormal{d}\mu\left(a\right)\,|\widehat{\left(A_{a}\right)}_{l0}|^{2}<\infty.\label{eq:cw12}\end{equation}

\subsection{Stereographic dilation}

\label{sub:Stereographic-dilation}In the original set up proposed
by \citet{antoine99} the stereographic dilation of functions is considered,
which is explicitly defined in real space on $\textnormal{S}^{2}$.
The stereographic dilation operator $D(a)$ on $G\in\textnormal{L}^{2}(\textnormal{S}^{2},\textnormal{d}\Omega)$,
for a continuous dilation factor $a\in\mathbb{R}_{+}^{*}$, is defined
in terms of the inverse of the corresponding stereographic dilation
$D_{a}$ on points in $\textnormal{S}^{2}$. It reads as

\begin{eqnarray}
G_{a}\left(\omega\right) & = & \left[D\left(a\right)G\right]\left(\omega\right)\nonumber \\
 & = & \lambda^{1/2}\left(a,\theta\right)G\left(D_{a}^{-1}\omega\right),\label{eq:cw13}\end{eqnarray}
with $\lambda^{1/2}(a,\theta)=a^{-1}[1+\tan^{2}(\theta/2)]/[1+a^{-2}\tan^{2}(\theta/2)]$.
The dilated point is given by $D_{a}(\theta,\varphi)=(\theta_{a}(\theta),\varphi)$
with the linear relation $\tan(\theta_{a}(\theta)/2)=a\tan(\theta/2)$.
The dilation operator therefore maps the sphere without its South
pole on itself: $\theta_{a}(\theta):\theta\in[0,\pi)\rightarrow\theta_{a}\in[0,\pi)$.
This dilation operator is uniquely defined by the requirement of the
following natural properties. The dilation of points on $\textnormal{S}^{2}$
must be a radial (\emph{i.e.} only affecting the radial variable $\theta$
independently of $\varphi$, and leaving $\varphi$ invariant) and
conformal (\emph{i.e.} preserving the measure of angles in the tangent
plane at each point) diffeomorphism (\emph{i.e.} a continuously differentiable
bijection). The normalization by $\lambda^{1/2}(a,\theta)$ in (\ref{eq:cw13})
is uniquely determined by the requirement that the dilation of functions
in $\textnormal{L}^{2}(\textnormal{S}^{2},\textnormal{d}\Omega)$
be a unitary operator (\emph{i.e.} preserving the scalar product in
$\textnormal{L}^{2}(\textnormal{S}^{2},\textnormal{d}\Omega)$, and
specifically the norm of functions). Notice that the stereographic
dilation operation is supported by a group structure for the composition
law of the corresponding operator $D(a)$. A group homomorphism also
holds with the operation of multiplication by $a$ on $\mathbb{R}_{+}^{*}$.

In this setting, the effect of the dilation on the spherical harmonic
coefficients of the dilated function is not easily tractable analytically.
Consequently, the admissibility condition (\ref{eq:cw8}) is difficult
to check in practice. On the contrary, wavelets on the plane are well-known,
and may be easily constructed, as the corresponding admissibility
condition reduces to a zero mean condition for a function both integrable
and square-integrable. In that context, a correspondence principle
was proved \citep{wiaux05}, stating that the inverse stereographic
projection of a wavelet on the plane leads to a wavelet on the sphere.
This correspondence principle notably requires the definition of a
scale integration measure identical to the measure used on the plane:
$\textnormal{d}\mu(a)=a^{-3}\textnormal{d}a$. Notice that this measure
naturally appears in the original group-theoretic context \citep{antoine99}.

\subsection{Harmonic dilation}

\label{sub:Harmonic-dilation}Another possible definition of the dilation
of functions may be considered, which is explicitly defined in harmonic
space on $\textnormal{S}^{2}$. It was proposed in previous developments
relative to the definition of a wavelet formalism on the sphere \citep{holschneider96,mcewen06}.
The harmonic dilation is defined directly on $G\in\textnormal{L}^{2}(\textnormal{S}^{2},d\Omega)$
through a sequence of prescriptions rather than in terms of the application
of an simple operator. Firstly, an arbitrary prescription must be
chosen to define a set of generating functions $\tilde{G}_{m}(k)$
of a continuous variable $k\in\mathbb{R}_{+}$, for each $m\in\mathbb{Z}$.
These functions are identified to the spherical harmonic coefficients
of $G$ through: $\tilde{G}_{m}(l)=\widehat{G}_{lm}$ for $l\in\mathbb{N}$,
and $|m|\leq l$. Secondly, the variable $k$ is dilated linearly,
$k=l\rightarrow k=al$, just as would be the norm of the Fourier frequency
on the plane. For a continuous dilation factor $a\in\mathbb{R}_{+}^{*}$,
the spherical harmonic coefficients of the dilated function $G_{a}$
are defined by: \begin{equation}
\widehat{\left(G_{a}\right)}_{lm}=\tilde{G}_{m}\left(al\right).\label{eq:cw14}\end{equation}

In the corresponding continuous wavelet formalism, the analysis function
$\Psi$ must satisfy the following form of the admissibility condition
(\ref{eq:cw8}). On the one hand $\widehat{\Psi}_{00}=\tilde{\Psi}_{0}\left(0\right)=0$,
which corresponds to the requirement that $\Psi$ has a zero mean
on the sphere: \begin{equation}
\frac{1}{4\pi}\int_{\textnormal{S}^{2}}\textnormal{d}\Omega\,\Psi\left(\omega\right)=0.\label{eq:cw15}\end{equation}
This zero mean is of course preserved through harmonic dilation. As
the zero frequency is not supported by the wavelets, only signals
with zero mean can be analyzed in this formalism (see relation (\ref{eq:cw6})).
Let us remark that wavelets on the sphere dilated through the stereographic
dilation do not necessarily have a zero mean. On the other hand, the
scale integration measure can arbitrarily be chosen as $\textnormal{d}\mu(a)=a^{-1}\textnormal{d}a$.
This leads to a simple expression of the remaining constraints for
$l\in\mathbb{N}^{0}$ as \begin{equation}
0<C_{\Psi}^{l}=\frac{8\pi^{2}}{2l+1}\sum_{|m|\leq l}\int_{\mathbb{R}_{+}}\frac{\textnormal{d}k'}{k'}\,|\tilde{\Psi}_{m}\left(k'\right)|^{2}<\infty.\label{eq:cw16}\end{equation}
The left-hand side inequality implies $0<\int_{\mathbb{R}_{+}}\textnormal{d}k'/k'\,|\tilde{\Psi}_{m_{0}}(k')|^{2}$
for at least one of the first two generating functions: $m_{0}\in\{0,1\}$.
In other words, either $\tilde{\Psi}_{0}$ or $\tilde{\Psi}_{1}$
must be non-zero on a set of non-zero measure on $\mathbb{R}_{+}$.
The right-hand side inequality implies $\int_{\mathbb{R}_{+}}\textnormal{d}k'/k'\,|\tilde{\Psi}_{m}(k')|^{2}<\infty$
for all generating functions: $m\in\mathbb{Z}$. Hence, the generating
functions must satisfy $\tilde{\Psi}_{m}(0)=0$ (this condition encompasses
the zero mean condition (\ref{eq:cw15}) in the form $\tilde{\Psi}_{0}(0)=0$)
and tend to zero when $k'\rightarrow\infty$. With this choice of
scale integration measure, the constraints summarize to the requirement
that each generating function satisfies a condition very similar to
the wavelet admissibility condition for an axisymmetric wavelet on
the plane \citep{antoine04}%
\footnote{The exact wavelet admissibility condition on the plane reduces to
a zero mean condition for functions that are both integrable and square-integrable.%
} defined by a Fourier transform identical to $\tilde{\Psi}_{m}\left(k\right)$.
Consequently, the wavelet admissibility condition (\ref{eq:cw16})
can be checked in practice and wavelets associated with the harmonic
dilation can be designed easily.

For continuous axisymmetric wavelets, a unique generating function
$\tilde{A}_{0}(k)$ of a continuous variable $k\in\mathbb{R}_{+}$
is required. The admissibility condition (\ref{eq:cw12}) reduces
to the following expression. The analysis function $A$ must have
a zero mean and only allows the analysis of signals with zero mean.
A unique additional condition holds independently of $l$: \begin{equation}
0<C_{A}=\int_{\mathbb{R}_{+}}\frac{\textnormal{d}k'}{k'}\,|\tilde{A}_{0}\left(k'\right)|^{2}<\infty.\label{eq:cw17}\end{equation}
This condition actually encompasses the zero mean condition in the
form $\tilde{A}_{0}(0)=0$, and also requires that the generating
function must tend to zero when $k'\rightarrow\infty$. The coefficients
entering the reconstruction formula (\ref{eq:cw11}) read as $C_{A}^{l}=4\pi C_{A}/(2l+1)$,
for $l\in\mathbb{N}^{0}$.

\subsection{Discussion}

On the one hand, the harmonic dilation lacks some of the important
properties which hold under stereographic dilation. As the harmonic
dilation does not act on points, the question of the corresponding
properties of a radial and conformal diffeomorphism make no sense.
The harmonic dilation of functions is not either a unitary procedure.
It does not preserve the scalar product in $\textnormal{L}^{2}(\textnormal{S}^{2},\textnormal{d}\Omega)$,
or specifically the norm of functions. This is due to the requirement
of definition of generating functions for any function to be dilated.
A group structure for the composition of harmonic dilations holds
only if successive dilations of a function $G$ are defined through
linear dilation of the variable $k$ of a unique generating function
$\tilde{G}_{m}(k)$ for each $m\in\mathbb{Z}$. The same condition
applies for the existence of a corresponding homomorphism structure
with the operation of multiplication by $a$ on $\mathbb{R}_{+}^{*}$.

Moreover, the harmonic dilation is explicitly defined in harmonic
space. The evolution in real space of localization and directionality
properties of functions on the sphere through harmonic dilation is
therefore not known analytically. However, in the Euclidean limit
where a function is localized on a small portion of the sphere, this
portion is assimilated to the tangent plane, and the stereographic
and harmonic dilations both identify with the standard dilation in
the plane. For each $m\in\mathbb{Z}$, the overall frequency index
$l\in\mathbb{N}\rightarrow\infty$ identifies with the continuous
variable $k\in\mathbb{R}_{+}\rightarrow\infty$, corresponding to
the norm of the Fourier frequency on the plane \citep{holschneider96}.
So in particular, the evolution of localization properties of functions
through harmonic dilation is at least controlled in the Euclidean
limit.

On the other hand, the very simple action of the harmonic dilation
in harmonic space also exhibits several advantages relative to the
stereographic dilation. Notably, the harmonic dilation ensures that
the band limit of a wavelet and of the corresponding wavelet coefficients,
is reduced by a factor $a$. Such a multi-resolution property is essential
in reducing the memory and computation time requirements for the wavelet
analysis of signals. It does not hold under stereographic dilation.
Moreover, as already emphasized, a scheme allowing an exact reconstruction
of signals from their wavelet coefficients requires a discretization
of the dilation factor. The definition of a scale discretized wavelet
formalism through an integration of the dilation factor $a$ by slices
in the continuous wavelet formalism turns out to be very natural with
a dilation defined in harmonic space, but not with the stereographic
dilation. Indeed, one would like the dilation operation acting on
scale discretized functions after the integration of the dilation
factor by slices to be the same as the original dilation operation.
It will become obvious that this property holds for a dilation defined
in harmonic space, but not for the stereographic dilation.

In conclusion, no obvious definition of dilation is imposed for the
development of a wavelet formalism on the sphere. But considering
our aim for a scale discretized wavelet formalism, as well as the
essential criterion of defining a formalism with multi-resolution
properties, we will focus on a scale discretized wavelet formalism
from a dilation defined in harmonic space. However, for any dilation
defined in harmonic space, the evolution of the localization and directionality
properties of functions in real space through dilation needs to be
understood and controlled. In that regard, we amend the harmonic dilation
(\ref{eq:cw14}) and define a kernel dilation to be applied on functions
which are said to be factorized steerable functions with compact harmonic
support. Moreover, the kernel dilation will also render the transition
between the continuous and scale discretized formalism much simpler
and more transparent than what the harmonic dilation can provide.

\section{Kernel dilation}

\label{sec:Kernel-dilation}In this section we define the kernel dilation
on factorized functions in harmonic space on the sphere. We consider
in particular factorized steerable functions with compact harmonic
support. We also study the localization and directionality properties
in real space for such functions, as well as the controlled evolution
of these properties through kernel dilation.

\subsection{Factorized functions and kernel dilation}

\label{sub:Factorized-functions}A function $G\in\textnormal{L}^{2}(\textnormal{S}^{2},\textnormal{d}\Omega)$
can be defined to be a factorized function in harmonic space if it
can be written in the form: \begin{equation}
\widehat{G}_{lm}=\tilde{K}_{G}\left(l\right)S_{lm}^{G},\label{eq:kd1}\end{equation}
for $l\in\mathbb{N}$, and $|m|\leq l$. The positive real \emph{kernel}
$\tilde{K}_{G}(k)\in\mathbb{R}_{+}$ is a generating function of a
continuous variable $k\in\mathbb{R}_{+}$, initially evaluated on
integer values $k=l$. The directionality coefficients $S_{lm}^{G}$
, for $l\in\mathbb{N}$, and $|m|\leq l$, define the \emph{directional
split} of the function. In particular, for a real function $G$, they
bear the same symmetry relation as the spherical harmonic coefficients
$\widehat{G}_{lm}$ themselves: $S_{lm}^{G*}=(-1)^{m}S_{l(-m)}^{G}$.
Without loss of generality one can impose \begin{equation}
\sum_{|m|\leq l}\vert S_{lm}^{G}\vert^{2}=1,\label{eq:kd2}\end{equation}
for the values of $l$ for which $S_{lm}^{G}$ is non-zero for at
least one value of $m$. Hence localization properties of a function
$G$, such as a measure of dispersion of angular distances around
its central position as weighted by the function values, are governed
by the kernel and to a lesser extent by the directional split. Indeed,
the power contained in the function $G$ at each allowed value of
$l$ is fixed by the kernel only. The norm of $G\in\textnormal{L}^{2}(\textnormal{S}^{2},\textnormal{d}\Omega)$
reads as $\vert\vert G\vert\vert^{2}=\sum_{l\in\mathbb{N}}\tilde{K}_{G}^{2}(l)$,
where the sum runs over the values of $l$ for which $S_{lm}^{G}$
is non-zero for at least one value of $m$. However, the directional
split is essential in defining the directionality properties measuring
the behaviour of the function with the azimuthal variable $\varphi$,
because of it bears the entire dependence of the spherical harmonic
coefficients of the function in the index $m$.

The kernel dilation applied to a factorized function (\ref{eq:kd1})
is simply defined by application of the harmonic dilation (\ref{eq:cw14})
to the kernel only. The directionality of the dilated function is
defined through the same directional split as the original function.
For a continuous dilation factor $a\in\mathbb{R}_{+}^{*}$, the dilated
function therefore reads as: \begin{equation}
\widehat{\left(G_{a}\right)}_{lm}=\tilde{K}_{G}\left(al\right)S_{lm}^{G}.\label{eq:kd3}\end{equation}
Let us emphasize that the directionality coefficients $S_{lm}^{G}$
are not affected by dilations, on the contrary of what the complete
action of the harmonic dilation (\ref{eq:cw14}) would imply. The
kernel and harmonic dilations strictly identify with one another when
applied to factorized axisymmetric functions $A$, for which the directional
split takes the trivial values $S_{lm}^{A}=\delta_{m0}$ for $l\in\mathbb{N}$.

\subsection{Compact harmonic support}

Any function $G\in\textnormal{L}^{2}(\textnormal{S}^{2},\textnormal{d}\Omega)$
can be said to have a compact harmonic support in the interval $l\in(\left\lfloor \alpha^{-1}B\right\rfloor ,B)$,
for any $B\in\mathbb{N}^{0}$ and any real value $\alpha>1$, if\begin{equation}
\widehat{G}_{lm}=0\quad\mbox{for\, all}\quad l,m\quad\mbox{with}\quad l\notin\left(\left\lfloor \alpha^{-1}B\right\rfloor ,B\right),\label{eq:kd4}\end{equation}
where $\left\lfloor x\right\rfloor $ denotes the largest integer
value below $x\in\mathbb{R}$. Notice that the compactness of the
harmonic support of $G$ can be defined as the ratio of the band limit
to the width of its support interval.

For a factorized function $G$ of the form (\ref{eq:kd1}), the compact
harmonic support in the interval $l\in(\left\lfloor \alpha^{-1}B\right\rfloor ,B)$
is ensured by the choice of a kernel with compact support in the interval
$k\in(\alpha^{-1}B,B)$:\begin{equation}
\tilde{K}_{G}\left(k\right)=0\quad\mbox{for}\quad k\notin(\alpha^{-1}B,B).\label{eq:kd5}\end{equation}
The compactness of the harmonic support of $G$ can simply be estimated
from the compact support of the kernel as $c(\alpha)=\alpha/(\alpha-1)\in[1,\infty)$.
One has $c(\alpha)\rightarrow\infty$ when $\alpha\rightarrow1$,
and $c(\alpha)\rightarrow1$ when $\alpha\rightarrow\infty$. Typical
values would be $\alpha=2$ corresponding to a compactness $c(2)=2$,
or $\alpha=1.1$ leading to a higher compactness $c(1.1)=11$.

By a kernel dilation with a dilation factor $a\in\mathbb{R}_{+}^{*}$
in (\ref{eq:kd3}), the compact support of the dilated kernel $\tilde{K}_{G}\left(ak\right)\in\mathbb{R}_{+}$
is defined in the interval $k\in(a^{-1}\alpha^{-1}B,a^{-1}B)$. The
compact harmonic support of the dilated function $G_{a}$ itself is
thus defined in the corresponding interval $l\in(\left\lfloor a^{-1}\alpha^{-1}B\right\rfloor ,\left\lceil a^{-1}B\right\rceil )$,
where $\left\lceil x\right\rceil $ denotes the smallest integer value
above $x\in\mathbb{R}$. In particular, the compactness of the harmonic
support of a function remains invariant through a kernel dilation.

\subsection{Steerable functions}

The notion of steerability was first introduced on the plane \citep{freeman91,simoncelli92},
and more recently defined on the sphere \citep{wiaux05}. By definition,
$G\in\textnormal{L}^{2}(\textnormal{S}^{2},\textnormal{d}\Omega)$
is steerable if any rotation of the function around itself may be
expressed as a linear combination of a finite number $M$ of basis
functions $G_{p}$: \begin{equation}
G_{\chi}\left(\omega\right)=\sum_{p=0}^{M-1}k_{p}\left(\chi\right)G_{p}\left(\omega\right).\label{eq:kd6}\end{equation}
The square-integrable functions $k_{p}(\chi)$ on the circle $\textnormal{S}^{1}$,
with $1\leq m\leq M$, and $M\in\mathbb{N}^{0}$, are called interpolation
weights. Intuitively, steerable functions have a non-zero angular
width in the azimuthal angle $\varphi$, which renders them sensitive
to a range of directions and enables them to satisfy the steerability
relation. This non-zero angular width naturally corresponds to an
azimuthal band limit $N\in\mathbb{N}^{0}$ in the frequency index
$m$ associated with the azimuthal variable $\varphi$: \begin{equation}
\widehat{G}_{lm}=0\quad\mbox{for\, all}\quad l,m\quad\mbox{with}\quad\vert m\vert\geq N.\label{eq:kd7}\end{equation}
It can actually be shown that the property of steerability (\ref{eq:kd6})
is equivalent to the existence of an azimuthal band limit $N$ (\ref{eq:kd7}).

On the one hand, if a function $G$ is steerable with $M$ basis functions,
then the number $T$ of values of $m$ for which $\widehat{G}_{lm}$
has a non-zero value for at least one value of $l$ is lower or equal
to $M$: $M\geq T$. This was firstly established for functions on
the plane \citep{freeman91}, and the proof is absolutely identical
on the sphere. As a consequence, the function has some azimuthal band
limit $N$, with $T\leq2N-1$.

On the other hand, if a function $G$ has an azimuthal band limit
$N$, then it is steerable, and the number of basis functions can
be reduced at least to $M=2N-1$. This second part of the equivalence
can be proved by explicitly deriving a steerability relation for band-limited
functions with an azimuthal band limit $N$. Any band-limited function
$G$ can in particular be steered using $M$ rotated versions $G_{\chi_{p}}=R^{\hat{z}}(\chi_{p})G$
as basis functions, and interpolation weights given by simple translations
by $\chi_{p}$ of a unique square-integrable function $k(\chi)$ on
the circle $\textnormal{S}^{1}$: \begin{equation}
G_{\chi}\left(\omega\right)=\sum_{p=0}^{M-1}k\left(\chi-\chi_{p}\right)G_{\chi_{p}}\left(\omega\right),\label{eq:kd8}\end{equation}
for specific rotation angles $\chi_{p}$ with $0\leq p\leq M-1$.
One may choose $M=2N-1$ equally spaced rotation angles $\chi_{p}\in[0,2\pi)$
as $\chi_{p}=2\pi p/(2N-1)$, with $0\leq p\leq2N-2$. The function
$k(\chi)$ is then defined by the Fourier coefficients $\widehat{k}_{m}=1/(2N-1)$
for $|m|\leq N-1$ and $\widehat{k}_{m}=0$ otherwise. Notice that
the angles $\chi_{p}$ and the structure of the function $k(\chi)$
are independent of the explicit non-zero values $\widehat{G}_{lm}$.

Typically, if $\widehat{G}_{lm}$ has a non-zero value for at least
one value of $l$ for all $m$ with $|m|\leq N-1$, then $T=2N-1$
and the function is optimally steered by these $M=T$ angles and the
function $k(\chi)$ described. On the contrary, when values of $m$,
with $|m|\leq N-1$, exist for which $\widehat{G}_{lm}=0$ for all
values of $l$, then $T<2N-1$ and one might want to reduce the number
$M=2N-1$ of basis functions. Depending on the distribution of the
$T$ values of $m$ for which $\widehat{G}_{lm}$ has a non-zero value
for at least one value of $l$, the number of basis functions required
to steer the band-limited function may indeed be optimized to its
smallest possible value $M=T$. This optimization is notably reachable
for functions with specific distributions of the $T$ values of $m$,
corresponding to particular symmetries in real space. For example,
a function $G$ is even or odd through rotation around itself by $\chi=\pi$
if and only if $\widehat{G}_{lm}$ has non-zero values only for, respectively,
even or odd values of $m$. This property notably implies that the
central position of the function $G$ identifies with the North pole,
in the sense that its modulus $|G|$ is then always even through rotation
around itself by $\chi=\pi$. The combination of an azimuthal band
limit $N$ with that symmetry reads as: \begin{equation}
\widehat{G}_{lm}=0\quad\mbox{for\, all}\quad l,m\quad\mbox{with}\quad m\notin T_{N},\label{eq:kd9}\end{equation}
with \begin{equation}
T_{N}=\left\{ -\left(N-1\right),-\left(N-3\right),...,\left(N-3\right),\left(N-1\right)\right\} .\label{eq:kd10}\end{equation}
In this particular case, $T=N$ and one may choose $M=N$ equally
spaced rotation angles $\chi_{p}\in[0,\pi)$ as $\chi_{p}=\pi p/N$,
with $0\leq p\leq N-1$, and steer the function through relation (\ref{eq:kd8}).
The function $k(\chi)$ is defined by the Fourier coefficients $\widehat{k}_{m}=1/N$
for $m\in T_{N}$ and $\widehat{k}_{m}=0$ otherwise.

In summary, the property of steerability is indeed equivalent to the
existence of an azimuthal band limit in $m$. For a factorized function
$G$ of the form (\ref{eq:kd1}), steerability constraints such as
(\ref{eq:kd7}) and (\ref{eq:kd9}) are ensured by the directionality
coefficients $S_{lm}^{G}$, independently of the kernel. Consequently,
any relation of steerability remains unchanged through a kernel dilation
(\ref{eq:kd3}), which by definition only affects the kernel.

\subsection{Localization control}

\label{sub:Localization-control}

Let us consider the Euclidean limit where a function is localized
on a small portion of the sphere which can be assimilated to the tangent
plane. As discussed, the harmonic dilation (\ref{eq:cw14}) identifies
with the standard dilation in the plane in that limit \citep{holschneider96}.
Hence, for factorized steerable functions with compact harmonic support,
the kernel dilation (\ref{eq:kd3}) certainly shares the same property
if it identifies with the harmonic dilation itself in the limit $l\rightarrow\infty$.
This is ensured by considering functions with directionality coefficients
$S_{lm}^{G}$ which become independent of $l$ in the limit $l\rightarrow\infty$.
Consequently, the evolution of localization properties of functions
through kernel dilation is also controlled in the Euclidean limit.
But a much more important localization property holds for the kernel
dilation at any frequency range for factorized functions with compact
harmonic support.

A typical localization property of a function $G\in\textnormal{L}^{2}(\textnormal{S}^{2},\textnormal{d}\Omega)$
is a measure of dispersion of angular distances around its central
position, as weighted by the function values. The corresponding measure
in harmonic space is defined by the dispersion of the values of $l$
around their central position, as weighted by the values of the spherical
harmonic coefficients $\widehat{G}_{lm}$, for each value of $m$.
It is well-known that the smaller the dispersion in real space, the
larger the dispersion in harmonic space. An optimal Dirac delta distribution
on the sphere $\delta_{\textnormal{S}^{2}}(\omega)$ exhibits an infinite
series in $l$ of spherical harmonic coefficients: $\widehat{(\delta_{\textnormal{S}^{2}})}_{lm}=[(2l+1)/4\pi]^{1/2}\delta_{m0}$.
On the contrary a spherical harmonic $Y_{lm}$, completely non-localized
in real space on $\textnormal{S}^{2}$, by definition exhibits a unique
frequency $l$.

In particular, we need to understand the evolution of this localization
property of a factorized steerable function $G$ with compact harmonic
support through the kernel dilation (\ref{eq:kd3}). Let us consider
for simplicity a factorized axisymmetric function $A$ with compact
harmonic support, for which the kernel and harmonic dilations identify
with one another. For an initial compact harmonic support in the interval
$l\in(\left\lfloor \alpha^{-1}B\right\rfloor ,B)$, the kernel dilation
by a factor $a$ modifies the interval to $l\in(\left\lfloor a^{-1}\alpha^{-1}B\right\rfloor ,\left\lceil a^{-1}B\right\rceil )$.
Hence in harmonic space the width of the harmonic support interval
is multiplied by $a^{-1}$. This also measures the evolution of the
dispersion in harmonic space. In real space, one can intuitively consider
that the corresponding dispersion of the values of the angular distance
$\theta$ around the North pole (which is the central position of
any axisymmetric function) is multiplied by $a$. This intuition is
actually only exact in the Euclidean limit $l\rightarrow\infty$,
reached when $a\rightarrow0$. But a weaker property holds though,
on a wide class of axisymmetric functions on the sphere, in particular
on factorized axisymmetric functions with compact harmonic support.
It takes the form of the following upper bound through the kernel
dilation by $a$ of such an axisymmetric function $A$ at a given
angular distance $\theta$ from the North pole:\begin{equation}
|A_{a}\left(\theta\right)|\leq b_{(A,k)}\frac{a^{-2}}{1+\left(\theta/a\right)^{k}},\label{eq:kd13}\end{equation}
for any integer $k\geq2$ and for some constant $b_{(A,k)}$ depending
on $A$ and $k$ \citep{narcowich05}. The ratio of the bounds at
the North pole and at any fixed angular distance $\theta$ simply
reads as $1+(\theta/a)^{k}$. When $a$ increases, this ratio gets
closer to unity and the bound is less constraining, enabling a larger
dispersion of the values of the angular distance $\theta$ around
the North pole. When $a$ decreases, the ratio increases and the bound
is more constraining, hence imposing a smaller dispersion of the values
of the angular distance $\theta$ around the North pole. This ensures
a good behaviour in real space for the kernel dilation, when applied
to factorized axisymmetric functions with compact harmonic support.

In summary, the dispersion of angular distances around the central
position of a function $G$ defines a localization property. We have
shown that the evolution of the localization of factorized axisymmetric
functions with compact harmonic support through kernel dilation is
controlled by the bound (\ref{eq:kd13}). For completeness, the corresponding
bound should be analyzed for the kernel dilation of factorized steerable
functions with compact harmonic support, but this goes beyond the
scope of the present work. The verification of more detailed localization
properties in real space for a function designed from its spherical
harmonic coefficients requires a numerical evaluation of sampled values
of that function.

\subsection{Directionality control}

\label{sub:Directionality-control}Let us consider a typical directionality
property of a function $G\in\textnormal{L}^{2}(\textnormal{S}^{2},\textnormal{d}\Omega)$,
such as measured by its auto-correlation function. The auto-correlation
function of $G$ is defined as the scalar product between two rotated
versions of the function around itself by angles $\chi,\chi'\in[0,2\pi)$.
This auto-correlation only depends on the difference of the rotation
angles $\Delta\chi=\chi-\chi'$ and is therefore considered in the
space $\textnormal{L}^{2}(\textnormal{S}^{1},\textnormal{d}\chi)$
of square-integrable functions on the circle $\textnormal{S}^{1}$:
$C^{G}(\Delta\chi)=\langle G_{\chi}\vert G_{\chi'}\rangle$. The peakedness
of the auto-correlation function in $\Delta\chi$ can be considered
as a measure of the directionality of the function: the more peaked
the auto-correlation, the more directional the function \citep{wiaux05}.
From the Plancherel relation $\langle F_{2}\vert F_{1}\rangle=\sum_{l\in\mathbb{N}}\sum_{|m|\leq l}\widehat{(F_{2})}_{lm}^{*}\widehat{(F_{1})}_{lm}$
for $F_{1},F_{2}\in\textnormal{L}^{2}(\textnormal{S}^{2},\textnormal{d}\Omega)$,
and the expression $\widehat{(G_{\chi})}_{lm}=e^{-im\chi}\widehat{G}_{lm}$
for the action of the operator $R^{\hat{z}}(\chi)$ on $G$, one gets
\begin{equation}
C^{G}\left(\Delta\chi\right)=\sum_{l\in\mathbb{N}}\sum_{|m|\leq l}e^{-im\Delta\chi}\vert\widehat{G}_{lm}\vert^{2}.\label{eq:kd14}\end{equation}
The value of the auto-correlation function at $\Delta\chi=0$ obviously
defines the square of the norm of the function : $C^{G}\left(0\right)=\vert\vert G\vert\vert^{2}$.

For a factorized function (\ref{eq:kd1}), the auto-correlation function
is strongly related to the directional split. Let us also recall that
in the case of a steerable function $G$ defined through (\ref{eq:kd8}),
the interpolation weights depend on the values of $m$ for which the
spherical harmonic coefficients have non-zero values and on the rotation
angles $\chi_{p}$, but not on the values $\widehat{G}_{lm}$ themselves.
This leaves enough freedom to design a suitable auto-correlation function
and thus control the directionality of the function. Let us also consider
a compact harmonic support (\ref{eq:kd4}) in the interval $(\left\lfloor \alpha^{-1}B\right\rfloor ,B)$.
We analyze the particular case where the directionality coefficients
are independent of $l$ for $l\geq N-1$,\begin{equation}
S_{lm}^{G}=S_{(N-1)m}^{G}\quad\mbox{for\, all}\quad l,m\quad\mbox{with}\quad l\geq N-1,\label{eq:kd15}\end{equation}
and where $N-1$ is lower or equal to the lowest integer value above
the lower bound of the compact harmonic support interval, \emph{i.e.}
$N-1\leq\left\lfloor \alpha^{-1}B\right\rfloor +1$. In that limit,
the auto-correlation reads as \begin{equation}
C^{G}\left(\Delta\chi\right)=\vert\vert G\vert\vert^{2}\sum_{|m|\leq N-1}e^{-im\Delta\chi}\vert S_{(N-1)m}^{G}\vert^{2}.\label{eq:kd16}\end{equation}
In other words, the square of the complex norm of the directionality
coefficients identifies with the Fourier coefficients of $C^{G}(\Delta\chi)$
in $\textnormal{L}^{2}(\textnormal{S}^{1},\textnormal{d}\chi)$. Notice
that a better directionality of a steerable function, as measured
by its auto-correlation function, is inevitably associated with a
larger band limit $N$, and with a larger number $T$ of values of
$m$ for which $\widehat{G}_{lm}$ has a non-zero value for at least
one value of $l$. Indeed, on the circle $S^{1}$ as on the plane
or the sphere, the smaller the dispersion of $\Delta\chi$ in real
space, the larger the dispersion of $m$ in harmonic space. Consequently,
a better directionality of a steerable function requires an increased
number $M$ of basis functions.

We need to understand the evolution of this directionality property
of a factorized steerable function $G$ with compact harmonic support
through the kernel dilation (\ref{eq:kd3}). The correlation function
of two dilated versions $G_{a}$ and $G_{a'}$ by factors $a$ and
$a'$ in $\mathbb{R}_{+}^{*}$ is defined through the scalar product
$C_{aa'}^{G}(\Delta\chi)=\langle G_{\chi,a}\vert G_{\chi',a'}\rangle$.
We consider again the case where the directionality coefficients are
independent of $l$ for $l\geq N-1$ and where the azimuthal band
limit for steerability is lower than the lower bound of the compact
harmonic support of each of the two dilated versions of $G$: $N-1\leq\left\lfloor a^{-1}\alpha^{-1}B\right\rfloor +1$
and $N-1\leq\left\lfloor a'^{-1}\alpha^{-1}B\right\rfloor +1$. In
that limit, the correlation $C_{aa'}^{G}\left(\Delta\chi\right)$
reads as \begin{equation}
C_{aa'}^{G}\left(\Delta\chi\right)=\langle G_{a}\vert G_{a'}\rangle\sum_{|m|\leq N-1}e^{-im\Delta\chi}\vert S_{(N-1)m}^{G}\vert^{2},\label{eq:kd17}\end{equation}
and appears to be simply proportional to the auto-correlation function
of $G$. For $a=a'$, this result states that the auto-correlations
of $G$ and $G_{a}$ are proportional. The control of directionality
of $G$ through the auto-correlation function is therefore preserved
through kernel dilation. For $a\neq a'$, this result essentially
ensures that the kernel dilation does not introduce any unexpected
distortion in the shape of the function in real space on $\textnormal{S}^{2}$.

In addition to the auto-correlation function, symmetry properties
may also be imposed on the spherical harmonic coefficients $\widehat{G}_{lm}$,
which translate into simple directionality properties in real space
for $G$. Firstly, in the framework of a wavelet analysis on the sphere,
one generally imposes the symmetry relation $\widehat{G}_{lm}^{*}=(-1)^{m}\widehat{G}_{l(-m)}$
in order to restrict to real analysis functions: $G(\theta,\varphi)\in\mathbb{R}$.
Secondly, the constraint that $\widehat{G}_{lm}$ has non-zero values
only for even or odd values $m\in T_{N}$, for any azimuthal band
limit $N$, implies that the function $G$ is respectively even or
odd under a rotation around itself by $\chi=\pi$: $G(\theta,\varphi+\pi)=(-1)^{N-1}G(\theta,\varphi)$.
Thirdly, for such functions, the additional constraint that the spherical
harmonic coefficients $\widehat{G}_{lm}$ are real for even values
of $N-1$, and purely imaginary for odd values of $N-1$, implies
that the function $G$ is respectively even or odd under a change
of sign on $\varphi$: $G(\theta,-\varphi)=(-1)^{N-1}G(\theta,\varphi)$.
These symmetries are defined up to a rotation of the function around
itself by any angle $\chi\in[0,2\pi)$, which amounts to a multiplication
of the spherical harmonic coefficients $\widehat{G}_{lm}$ by a complex
phase $e^{-im\chi}$. The three properties discussed are obviously
preserved through kernel dilation of factorized functions. They indeed
only concern the directionality coefficients $S_{lm}^{G}$, $ $which
are not affected by the kernel dilation.

In summary, the auto-correlation function and additional symmetries
define directionality properties of a function. We have shown that
the directionality properties studied are essentially preserved through
kernel dilation of factorized steerable functions with compact harmonic
support. Again, the verification of more precise directionality properties
in real space for a function designed from its spherical harmonic
coefficients unavoidably requires a numerical evaluation of sampled
values of that function.

\section{Wavelets from kernel dilation}

\label{sec:Scale-discretized-wavelets}In this section we begin with
the derivation of a new continuous wavelet formalism from the kernel
dilation with continuous scales, and for factorized steerable wavelets
with compact harmonic support. We then derive the scale discretized
wavelet formalism from the continuous wavelet formalism. The transition
is performed through an integration of the dilation factor by slices.
We emphasize the practical accessibility of an exact reconstruction
of band-limited signals from a finite number of analysis scales. We
also illustrate these developments through the explicit design of
an example scale discretized wavelet. We finally recast the scale
discretized wavelet formalism developed in a generic invertible filter
bank perspective.

\subsection{Continuous wavelets}

\label{sub:Suitable-wavelet-family}We simply consider the continuous
wavelet formalism exposed in Section \ref{sec:Continuous-wavelets},
and particularize it to the kernel dilation defined in Section \ref{sec:Kernel-dilation}.
Hence the scales of analysis are still continuous. The translations
by $\omega_{0}\in\textnormal{S}^{2}$ and proper rotations by $\chi\in[0,2\pi)$
of the wavelets are still defined through the continuous three-dimensional
rotations from relation (\ref{eq:cw2}) and (\ref{eq:cw3}).

For application of the kernel dilation, we consider continuous factorized
steerable functions $\Psi\in\textnormal{L}^{2}(\textnormal{S}^{2},\textnormal{d}\Omega)$
with compact harmonic support: \begin{equation}
\widehat{\Psi}_{lm}=\tilde{K}_{\Psi}\left(l\right)S_{lm}^{\Psi},\label{eq:kd18}\end{equation}
for a continuous kernel defined by a positive real function $\tilde{K}_{\Psi}(k)\in\mathbb{R}_{+}$
and a directional split defined by the directionality coefficients
$S_{lm}^{\Psi}$. The compact harmonic support of the wavelet in the
interval $l\in(\left\lfloor \alpha^{-1}B\right\rfloor ,B)$ is ensured
by a kernel $\tilde{K}_{\Psi}(k)$ with compact support in the interval
$k\in(\alpha^{-1}B,B)$, with a compactness $c(\alpha)=\alpha/(\alpha-1)\in[1,\infty)$:\begin{equation}
\tilde{K}_{\Psi}\left(k\right)=0\quad\mbox{for}\quad k\notin(\alpha^{-1}B,B).\label{eq:kd19}\end{equation}
The steerability of a wavelet with an azimuthal band limit $N$ in
ensured by the directional split: \begin{equation}
S_{lm}^{\Psi}=0\quad\mbox{for\, all}\quad l,m\quad\mbox{with}\quad\vert m\vert\geq N,\label{eq:kd20}\end{equation}
with\begin{equation}
\sum_{|m|\leq\min\left(N-1,l\right)}\vert S_{lm}^{\Psi}\vert^{2}=1,\label{eq:kd21}\end{equation}
for all $l\in\mathbb{N}^{0}$. Continuous axisymmetric wavelets $A(\theta)$
with compact harmonic support are simply obtained by the trivial directional
split with $S_{lm}^{A}=\delta_{m0}$ for all $l\in\mathbb{N}^{0}$.

The analysis of a signal $F\in\textnormal{L}^{2}(\textnormal{S}^{2},\textnormal{d}\Omega)$
with the analysis function $\Psi$ gives the wavelet coefficients
$W_{\Psi}^{F}\left(\rho,a\right)$ at each continuous scale $a$,
around each point $\omega_{0}$, and in each orientation $\chi$,
through the directional correlation (\ref{eq:cw4}). The reconstruction
of $F$ from its wavelet coefficients results from relation (\ref{eq:cw7}).
The zero mean condition (\ref{eq:cw15}) for the admissibility of
$\Psi$ implies $\tilde{K}_{\Psi}^{2}(0)=0$. One can also set arbitrarily
$S_{00}^{\Psi}=0$. The admissibility condition (\ref{eq:cw16}) summarizes
to: \begin{equation}
0<C_{\Psi}=\int_{(\alpha^{-1}B,B)}\frac{\textnormal{d}k'}{k'}\,\tilde{K}_{\Psi}^{2}\left(k'\right)<\infty,\label{eq:kd22}\end{equation}
which actually also encompasses the zero mean condition. The coefficients
entering the reconstruction formula are $C_{\Psi}^{l}=8\pi^{2}C_{\Psi}/(2l+1)$
for $l\in\mathbb{N}^{0}$. In other words, the kernel must formally
be identified with the Fourier transform of an axisymmetric wavelet
on the plane.

Notice that for a factorized wavelet $\Psi$, the directional correlation
defining the analysis of a signal may also be understood as a double
correlation, by the kernel and the directional split successively.
The standard correlation (\ref{eq:cw9}) of the signal $F$ and the
axisymmetric wavelets defined by the kernel of $\Psi$, provides intermediate
wavelet coefficients $W_{\tilde{K}_{\Psi}}^{F}(\omega_{0},a)$ on
$\textnormal{S}^{2}$ at each scale $a\in\mathbb{R}_{+}^{*}$. The
spherical harmonic transform of these coefficients reads as:\begin{equation}
\widehat{\left(W_{\tilde{K}_{\Psi}}^{F}\right)}_{lm}\left(a\right)=\sqrt{\frac{4\pi}{2l+1}}\tilde{K}_{\Psi}\left(al\right)\widehat{F}_{lm}.\label{eq:kd23}\end{equation}
At each scale $a$, the directional correlation of the intermediate
signal obtained at that scale $W_{\tilde{K}_{\Psi}}^{F}(\omega_{0},a)$
and a directional wavelet defined by the directional split of $\Psi$
provides the final wavelet coefficients on $\textnormal{SO(3)}$:
\begin{equation}
\widehat{\left(W_{\Psi}^{F}\right)}_{mn}^{l}\left(a\right)=\frac{8\pi^{2}}{2l+1}\left(\sqrt{\frac{2l+1}{4\pi}}S_{ln}^{\Psi}\right)^{*}\widehat{\left(W_{\tilde{K}_{\Psi}}^{F}\right)}_{lm}\left(a\right).\label{eq:kd24}\end{equation}
This reasoning obviously holds independently of the steerability or
compact harmonic support properties of $\Psi$.

In conclusion, the definition of the kernel dilation provides a new
continuous wavelet formalism, where scales, translations, and proper
rotations of the wavelets are all continuous. As the previously developed
continuous wavelet formalism based on the stereographic dilation,
it finds application in the identification of local directional features
of signals on the sphere. The wavelets defined bear new properties
of compact harmonic support and steerability, which are preserved
through kernel dilation. These properties can give a new insight for
the analysis of local directional features. However, as already discussed
the continuous scales required for the analysis prevent in practice
the exact reconstruction of the signals analyzed from their wavelet
coefficients.

\subsection{Scale discretized wavelets}

\label{sub:Scale-integration-by}Scale discretized wavelets $\Gamma$
can simply be obtained from continuous wavelets through an integration
by slices of the dilation factor $a\in\mathbb{R}_{+}^{*}$. Through
this transition procedure, scale discretized wavelets remain factorized
steerable functions with compact harmonic support, and are dilated
through kernel dilation.

We consider the analysis of a signal $F\in\textnormal{L}^{2}(\textnormal{S}^{2},\textnormal{d}\Omega)$
with band limit $B$. The original continuous wavelet $\Psi\in\textnormal{L}^{2}(\textnormal{S}^{2},\textnormal{d}\Omega)$
with a compact support is defined in the interval $k\in(\alpha^{-1}B,B)$.
The value $\alpha>1$ regulates the compactness $c(\alpha)$ of $\Psi$.
It is also taken as a basis dilation factor. The discrete dilation
factors for the scale discretized wavelet will correspond to integer
powers $\alpha^{j}$, for analysis depths $j\in\mathbb{N}$.

The scale discretized wavelet $\Gamma\in\textnormal{L}^{2}(\textnormal{S}^{2},\textnormal{d}\Omega)$
is thus defined in factorized form: \begin{equation}
\widehat{\Gamma}_{lm}=\tilde{K}_{\Gamma}\left(l\right)S_{lm}^{\Gamma},\label{eq:sdw1}\end{equation}
for a scale discretized kernel defined by a positive real function
$\tilde{K}_{\Gamma}(k)\in\mathbb{R}_{+}$ and a directional split
defined by the directionality coefficients $ $$S_{lm}^{\Gamma}$.
The directional split of $\Gamma$ is identified with the split of
$\Psi$:\begin{equation}
S_{lm}^{\Gamma}=S_{lm}^{\Psi},\label{eq:sdw2}\end{equation}
also giving\begin{equation}
S_{lm}^{\Gamma}=0\quad\mbox{for\, all}\quad l,m\quad\mbox{with}\quad\vert m\vert\geq N,\label{eq:sdw3}\end{equation}
and \begin{equation}
\sum_{|m|\leq\min\left(N-1,l\right)}\vert S_{lm}^{\Gamma}\vert^{2}=1,\label{eq:sdw4}\end{equation}
 for $l\in\mathbb{N}^{0}$, while $S_{00}^{\Gamma}=0$. The exact
same steerability properties are therefore obviously shared by the
continuous wavelet and the scale discretized wavelet, independently
of any dilation factor. The scale discretized kernel $\tilde{K}_{\Gamma}(k)$
is obtained from the continuous kernel $\tilde{K}_{\Psi}(k)$ through
an integration by slices of the dilation factor $a\in\mathbb{R}_{+}^{*}$
of the continuous wavelet formalism.

As a first step, a positive real \emph{scaling function} $\tilde{\Phi}_{\Gamma}(k)\in\mathbb{R}_{+}$
of a continuous variable $k\in\mathbb{R}_{+}$, is defined which gathers
the largest dilation factors $a\in(1,\infty)$, or correspondingly
the lowest values of $k$. This generating function reads for $k\in\mathbb{R}_{+}^{*}$
as:\begin{eqnarray}
\tilde{\Phi}_{\Gamma}^{2}\left(k\right) & = & \frac{1}{C_{\Psi}}\int_{1}^{\infty}\frac{\textnormal{d}a}{a}\,\tilde{K}_{\Psi}^{2}\left(ak\right)\nonumber \\
 & = & \frac{1}{C_{\Psi}}\int_{(\alpha^{-1}B,B)\cap(k,\infty)}\frac{\textnormal{d}k'}{k'}\,\tilde{K}_{\Psi}^{2}\left(k'\right),\label{eq:sdw5}\end{eqnarray}
and continuously continuated at $k=0$ by $\tilde{\Phi}_{\Gamma}^{2}(k)=1$.
The scaling function $\tilde{\Phi}_{\Gamma}^{2}\left(k\right)$ therefore
decreases continuously from unity down to zero in the interval $k\in(\alpha^{-1}B,B)$:\begin{eqnarray}
\tilde{\Phi}_{\Gamma}^{2}\left(k\right) & = & 1\quad\mbox{for}\quad0\leq k\leq\alpha^{-1}B,\nonumber \\
\tilde{\Phi}_{\Gamma}^{2}\left(k\right) & \in & (0,1)\quad\mbox{for}\quad\alpha^{-1}B<k<B,\nonumber \\
\tilde{\Phi}_{\Gamma}^{2}\left(k\right) & = & 0\quad\mbox{for}\quad k\geq B.\label{eq:sdw6}\end{eqnarray}
Notice that similar procedures of scale integration by slices were
already proposed in the development of corresponding formalisms on
the plane \citep{duval93,muschietti95,vandergheynst02}.

As a second step, a simple Littlewood-Paley decomposition \citep{frazier91}
is used to define the scale discretized kernel $\tilde{K}_{\Gamma}(k)$
by subtracting the scaling function $\tilde{\Phi}_{\Gamma}(k)$ to
its contracted version $\tilde{\Phi}_{\Gamma}(\alpha^{-1}k)$. This
implicitly sets the value $\alpha$ as the basis dilation factor.
The scale discretized kernel also reads as an integration of the continuous
kernel over a slice $a\in(\alpha^{-1},1)$ for the dilation factor,
or equivalently over a slice $k\in(\alpha^{-1}B,B)\cap(\alpha^{-1}k,k)$
of the compact support interval:\begin{eqnarray}
\tilde{K}_{\Gamma}^{2}\left(k\right) & = & \tilde{\Phi}_{\Gamma}^{2}\left(\alpha^{-1}k\right)-\tilde{\Phi}_{\Gamma}^{2}\left(k\right)\nonumber \\
 & = & \frac{1}{C_{\Psi}}\int_{^{\alpha^{-1}}}^{1}\frac{\textnormal{d}a}{a}\,\tilde{K}_{\Psi}^{2}\left(ak\right)\nonumber \\
 & = & \frac{1}{C_{\Psi}}\int_{(\alpha^{-1}B,B)\cap(\alpha^{-1}k,k)}\frac{\textnormal{d}k'}{k'}\,\tilde{K}_{\Psi}^{2}\left(k'\right).\label{eq:sdw7}\end{eqnarray}
The scale discretized kernel therefore has a compact support in the
interval $k\in(\alpha^{-1}B,\alpha B)$: \begin{equation}
\tilde{K}_{\Gamma}\left(k\right)=0\quad\mbox{for}\quad k\notin\left(\alpha^{-1}B,\alpha B\right).\label{eq:sdw8}\end{equation}
This support is wider than for the original continuous kernel and
the scaling function. The corresponding compactness reads as $c(\alpha^{2})=\alpha^{2}/(\alpha^{2}-1)\in[1,\infty)$.
The compact harmonic support of the scale discretized wavelet $\Gamma$
itself is thus defined in the interval $l\in(\left\lfloor \alpha^{-1}B\right\rfloor ,\left\lceil \alpha B\right\rceil )$.
The kernel also satisfies $\tilde{K}_{\Gamma}^{2}(0)=0$, leading
to a scale discretized wavelet $\Gamma$ with a zero mean on the sphere:
\begin{equation}
\frac{1}{4\pi}\int_{\textnormal{S}^{2}}\textnormal{d}\Omega\,\Gamma\left(\omega\right)=0.\label{eq:sdw9}\end{equation}

The dilations by $\alpha^{j}$ of the scale discretized wavelet obtained
are defined by the kernels $\tilde{K}_{\Gamma}(\alpha^{j}k)$ for
any analysis depth $j\in\mathbb{N}$. Each kernel has a compact support
in the interval $k\in(\alpha^{-(1+j)}B,\alpha^{(1-j)}B)$ and exhibits
a maximum in $k=\alpha^{-j}B$, with $\tilde{K}_{\Gamma_{\alpha^{j}}}(\alpha^{-j}B)=1$.
The scale discretized wavelet $\Gamma_{\alpha^{j}}$ at each analysis
depth $j$ thus has a compact harmonic support in the interval $l\in(\left\lfloor \alpha^{-(1+j)}B\right\rfloor ,\left\lceil \alpha^{(1-j)}B\right\rceil )$.
The property $\tilde{K}_{\Gamma}^{2}(0)=0$ still ensures that each
scale discretized wavelet has a zero mean on the sphere. Notice that
for $j\geq1$, one gets a dilation factor strictly greater than unity
$\alpha^{j}>1$, and the scale discretized wavelet has a band limit
lower or equal to the assumed band limit $B$ for the signal $F$
to be analyzed. At $j=0$, only the values of the kernel in the interval
$l\in(\left\lfloor \alpha^{-1}B\right\rfloor ,B)$ are of interest,
as higher frequencies $l$ are truncated by the signal $F$ itself
through the directional correlation. One can equivalently consider
that the compact support of the kernel is restricted to $k\in(\alpha^{-1}B,B)$
in the definition of the scale discretized wavelet at this first analysis
depth $j=0$. For $j\leq-1$, the lower bound of the compact harmonic
support of the scale discretized wavelet is larger than the band limit
$B$. The scale discretized wavelets with negative analysis depths
can therefore be discarded, as the result of their directional correlation
with the signal $F$ would be identically zero.

The admissibility condition (\ref{eq:kd22}) for continuous wavelets
simply turns into a resolution of the identity below the band limit
by a set of dilated wavelets at various analysis depths $j$, with
$0\leq j\leq J$, and a dilated scaling function at some total analysis
depth $J\in\mathbb{N}$. One gets in particular for $0\leq k=l<B$:\begin{equation}
\tilde{\Phi}_{\Gamma}^{2}\left(\alpha^{J}l\right)+\sum_{j=0}^{J}\tilde{K}_{\Gamma}^{2}\left(\alpha^{j}l\right)=1.\label{eq:sdw10}\end{equation}
The scaling function values $\tilde{\Phi}_{\Gamma}(\alpha^{J}l)$
are equal to unity in the interval $l\in[0,\left\lfloor \alpha^{-(1+J)}B\right\rfloor ]$,
then decrease in the interval $l\in(\left\lfloor \alpha^{-(1+J)}B\right\rfloor ,\left\lceil \alpha^{-J}B\right\rceil )$,
and are equal to zero for $l\geq\left\lceil \alpha^{-J}B\right\rceil $.
The kernel values $\tilde{K}_{\Gamma}(\alpha^{j}l)$ are non-zero
only in the compact harmonic support interval $l\in(\left\lfloor \alpha^{-(1+j)}B\right\rfloor ,\left\lceil \alpha^{(1-j)}B\right\rceil )$.
The scaling function typically retains the low frequency part of the
signal, which will not be analyzed. All signal information at frequencies
$l\leq\left\lfloor \alpha^{-(1+J)}B\right\rfloor $ is kept only in
the scaling function, equal to unity. The wavelets are equal to zero
at these frequencies. All signal information at frequencies $l\geq\left\lceil \alpha^{-J}B\right\rceil $
is fully analyzed by the wavelets, while the scaling function is equal
to zero. Intermediate frequencies are also analyzed by the wavelets
but the scaling function is required for the reconstruction of the
corresponding signal information.

Let us define the maximum analysis depth $J_{B}(\alpha)$ as the lowest
integer value such that $\alpha^{-J_{B}(\alpha)}B\leq1$: \begin{equation}
J_{B}\left(\alpha\right)=\left\lceil \log_{\alpha}B\right\rceil .\label{eq:sdw11}\end{equation}
In a case where the total analysis depth would be chosen strictly
above $J_{B}(\alpha)$, all wavelets at analysis depths $j$ with
$J\geq j\geq J_{B}(\alpha)+1$ would be identically null as their
kernel have a compact support strictly included in the interval $k\in(0,1)$.
The total analysis depth is consequently naturally limited by $J\leq J_{B}(\alpha)$.
In the case $J=J_{B}(\alpha)$, the dilated scaling function evaluated
at $\alpha^{J_{B}(\alpha)}l$ has a non-zero value only at $l=0$,
$\tilde{\Phi}_{\Gamma}^{2}(\alpha^{J_{B}(\alpha)}l)=\delta_{l0}$,
while all wavelets are equal to zero at $l=0$ as they have a zero
mean. Hence, the identity can be resolved with $J_{B}(\alpha)+1$
dilated wavelets and a trivial scaling function which simply retains
the spherical harmonic coefficient $\widehat{F}_{00}$ out of the
analysis, or equivalently the mean of the signal over the sphere.
One gets in particular for $0\leq k=l<B$: \begin{equation}
\delta_{l0}+\sum_{j=0}^{J_{B}(\alpha)}\tilde{K}_{\Gamma}^{2}\left(\alpha^{j}l\right)=1.\label{eq:sdw12}\end{equation}

\subsection{Analysis and exact reconstruction}

\label{sub:Exact-reconstruction}Following the scale discretization
defining the wavelets $\Gamma\in\textnormal{L}^{2}(\textnormal{S}^{2},\textnormal{d}\Omega)$,
a new scale discretized wavelet formalism is provided for the analysis
and the exact reconstruction of band-limited signals.

The analysis of a band-limited signal $F\in\textnormal{L}^{2}(\textnormal{S}^{2},\textnormal{d}\Omega)$
with band limit $B$, with a scale discretized wavelet $\Gamma$ is
performed by directional correlations just as in the continuous wavelet
formalism. The translations by $\omega_{0}\in\textnormal{S}^{2}$
and proper rotations by $\chi\in[0,2\pi)$ of the wavelets are still
defined through the continuous three-dimensional rotations from relation
(\ref{eq:cw2}) and (\ref{eq:cw3}). At each analysis depth $j$ with
$0\leq j\leq J\leq J_{B}(\alpha)$, the analysis is performed by directional
correlations of $F$ with the analysis functions $\Gamma_{\alpha^{j}}$
dilated through the kernel dilation by dilation factors $\alpha^{j}$:\begin{equation}
W_{\Gamma}^{F}\left(\rho,\alpha^{j}\right)=\langle\Gamma_{\rho,\alpha^{j}}|F\rangle.\label{eq:sdw13}\end{equation}
At each discrete scale $\alpha^{j}$, the wavelet coefficients $W_{\Gamma}^{F}(\rho,\alpha^{j})$
still identify a square-integrable function on $\textnormal{SO(3)}$,
and characterize the signal around each point $\omega_{0}$, and in
each orientation $\chi$. Once more, the direct Wigner $D$-function
transform of the wavelet coefficients is given as the pointwise product
of the spherical harmonic coefficients of the signal and the wavelet:\begin{equation}
\widehat{\left(W_{\Gamma}^{F}\right)}_{mn}^{l}\left(\alpha^{j}\right)=\frac{8\pi^{2}}{2l+1}\widehat{\left(\Gamma_{\alpha^{j}}\right)}_{ln}^{*}\widehat{F}_{lm}.\label{eq:sdw14}\end{equation}
Again, the factorization relation (\ref{eq:sdw1}) allows one to understand
the directional correlation (\ref{eq:sdw14}) as a double correlation,
by the kernel and the directional split successively.

The reconstruction of the band-limited signal $F$ from its wavelet
coefficients reads in terms of a summation on a finite number $J+1$
of discrete dilation factors:\begin{eqnarray}
F\left(\omega\right) & = & \left[\Phi_{\alpha^{J}}F\right]\left(\omega\right)+\nonumber \\
 &  & \sum_{j=0}^{J}\int_{\textnormal{SO(3)}}\textnormal{d}\rho\, W_{\Gamma}^{F}\left(\rho,\alpha^{j}\right)\left[R\left(\rho\right)L^{\textnormal{d}}\Gamma_{\alpha^{j}}\right]\left(\omega\right).\nonumber \\
 &  & \,\label{eq:sdw15}\end{eqnarray}
 The approximation $[\Phi_{\alpha^{J}}F](\omega)$ accounts for the
part of the signal retained in the scaling function $\tilde{\Phi}_{\Gamma}(\alpha^{J}l)$.
In a very similar way to the part of the signal analyzed by the wavelets,
it can be written as: \begin{equation}
\left[\Phi_{\alpha^{J}}F\right]\left(\omega\right)=2\pi\int_{\textnormal{S}^{2}}\textnormal{d}\Omega_{0}\, W_{\Phi}^{F}\left(\omega_{0},\alpha^{J}\right)\left[R\left(\omega_{0}\right)L^{\textnormal{d}}\Phi_{\alpha^{J}}\right]\left(\omega\right),\label{eq:sdw16}\end{equation}
with $W_{\Phi}^{F}(\omega_{0},\alpha^{J})=\langle\Phi_{\omega_{0},\alpha^{J}}|F\rangle$,
and for an axisymmetric function $\Phi\in\textnormal{L}^{2}(\textnormal{S}^{2},\textnormal{d}\Omega)$
defined by $\widehat{(\Phi_{\Gamma})}_{lm}=\tilde{\Phi}_{\Gamma}(l)\delta_{m0}$.
In the particular case where $J=J_{B}(\alpha)$, one gets $\widehat{(\Phi_{\Gamma})}_{lm}=\delta_{l0}\delta_{m0}$
and the approximation simply reduces to the mean of the signal over
the sphere: $[\Phi_{\alpha^{J_{B}(\alpha)}}F]=(4\pi)^{-1}\int_{\textnormal{S}^{2}}\textnormal{d}\Omega\, F(\omega)$.
The zero mean signal is completely analyzed by the scale discretized
wavelets. The operator $L^{\textnormal{d}}$ in $\textnormal{L}^{2}(\textnormal{S}^{2},\textnormal{d}\Omega)$
in the present scale discretized wavelet formalism is defined by the
following action on the spherical harmonic coefficients of functions:
$\widehat{L^{\textnormal{d}}G}_{lm}=(2l+1)\widehat{G}_{lm}/8\pi^{2}$.
This operator defining the scale discretized wavelets $L^{\textnormal{d}}\Gamma_{\alpha^{j}}$
used for reconstruction is independent of $\Gamma$, contrarily to
the operator $L_{\Psi}$ for continuous wavelets. This simply comes
from the fact that the scale discretized wavelets are, through their
definition (\ref{eq:sdw7}), normalized by $C_{\Psi}$.

Just as in the continuous wavelet formalism where the admissibility
condition (\ref{eq:kd22}) is required, the present reconstruction
formula holds if and only if the scale discretized wavelet satisfies
the constraints (\ref{eq:sdw4}), and (\ref{eq:sdw10}) or (\ref{eq:sdw12}).
These constraints are automatically satisfied by construction of the
scale discretized wavelets through the integration by slices. Again,
this corresponds to the requirement that the wavelet family as a whole,
including the scaling function, preserves the signal information at
each frequency $l\in\mathbb{N}$. 

Let us emphasize the fact that a finite number of discrete dilation
factors is required for the analysis and reconstruction of a band-limited
signal. Contrarily to the case of the continuous dilation factors,
this allows exact reconstruction of band-limited signals from relation
(\ref{eq:sdw15}). The translations and proper rotations of the wavelets
are still defined through the continuous three-dimensional rotations.
As discussed in Subsection \ref{sub:Directional-wavelets}, the exact
reconstruction is achieved only for suitable pixelizations of $\rho=(\varphi_{0},\theta_{0},\chi)$
which provide an exact quadrature rule for the numerical integration
of band-limited functions on $\textnormal{SO(3)}$. In the case of
non band-limited signals, an infinite number of negative analysis
depths $j\leq-1$ should be added for a complete analysis. This would
break the possibility of exact reconstruction. But in any case, no
exact quadrature rule exists on $\textnormal{SO(3)}$ for the numerical
integration of non band-limited functions, which already prevents
an exact numerical analysis.

Let us also remark that scale discretized axisymmetric wavelets with
compact harmonic support and dilated through kernel dilation were
recently introduced under the name of needlets \citep{baldi06,guilloux07,marinucci07}.
It is possible to show that the needlet coefficients of a wide class
of random signals on the sphere are uncorrelated in the asymptotic
limit of small scales, at any fixed angular distance on $\textnormal{S}^{2}$.
The scale discretized steerable wavelets with compact harmonic support,
thanks to their factorized form and to the choice of the kernel dilation,
are also good candidates for a directional extension of needlets.

\subsection{Example wavelet design}

\label{sub:Example-wavelet-design}%
\begin{figure*}
\begin{center}\includegraphics[width=16.8cm]{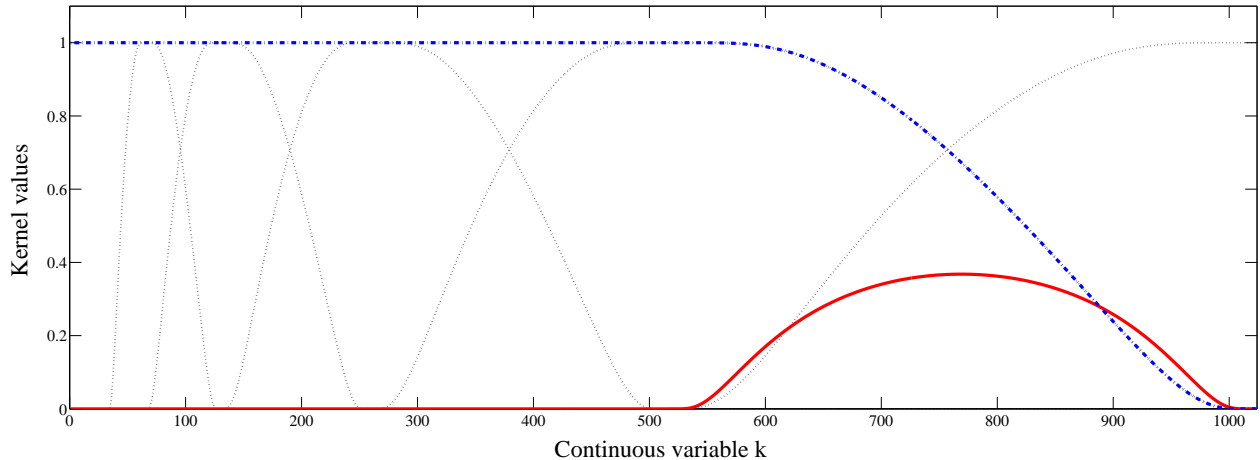}\end{center}

\caption{\label{fig:allkernels}Graphs of the continuous kernel defined in
(\ref{eq:sdw20}) and (\ref{eq:sdw21}), and the corresponding scale
discretized kernels obtained by differences of scaling functions at
various analysis depths. A band limit $B=1024$ and a basis dilation
factor $\alpha=2$ are chosen. The continuous kernel $\tilde{K}_{\Psi}(k)$
is represented by the continuous red line. The numerically integrated
scaling function $\tilde{\Phi}_{\Gamma}(k)$ is represented by the
dot-dashed blue line. The scale discretized kernels $\tilde{K}_{\Psi}(2^{j}k)$
are plotted as dotted black lines for the five first analysis depths
$j$, with $0\leq j\leq4$. For $j=0$, the corresponding compact
support interval is cut at the band limit: $k\in(512,1024)$. For
$1\leq j\leq4$ as for all larger analysis depths (not shown), the
intervals progressively move to lower frequencies and shrink: $k\in(256/2^{(j-1)},1024/2^{(j-1)})$.
At the maximum analysis depth $j=J_{B}(\alpha)=10$, the compact support
is shrunk to $k\in(0.5,2)$ and the scale discretized kernel only
contains the frequency $l=1$.}

\end{figure*}
\begin{figure*}
\begin{center}\includegraphics[width=4.2cm]{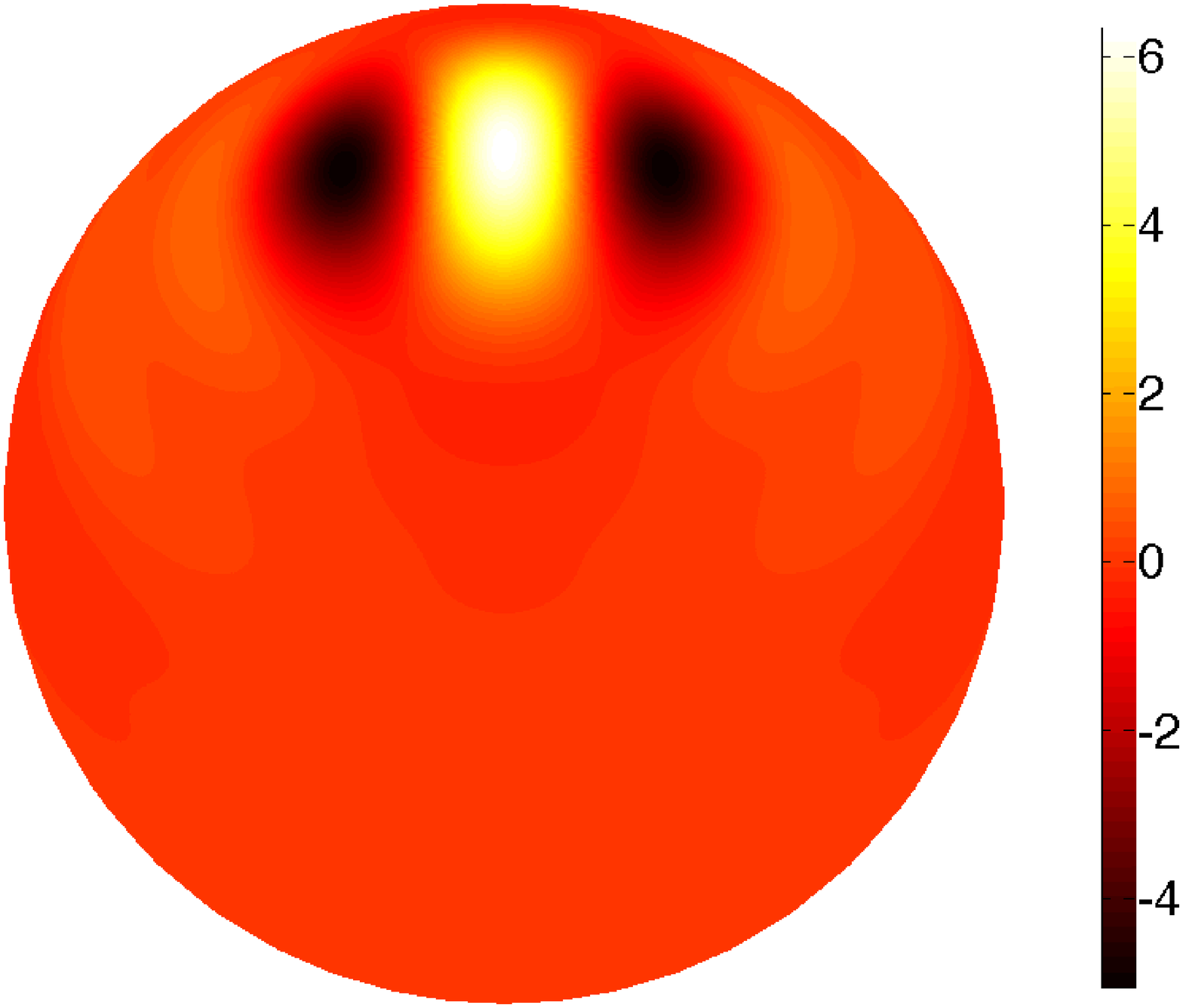}\includegraphics[width=4.2cm]{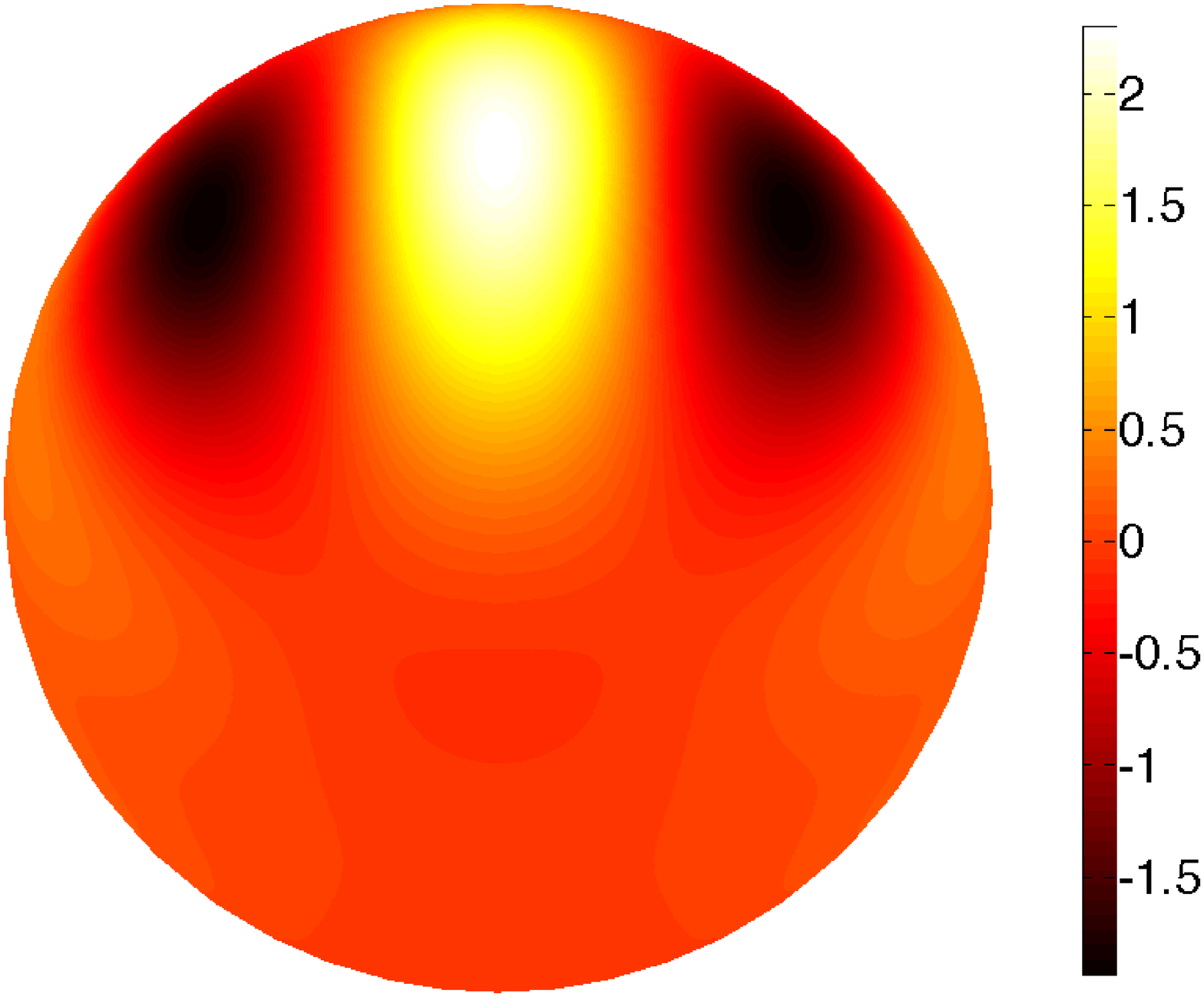}\includegraphics[width=4.2cm]{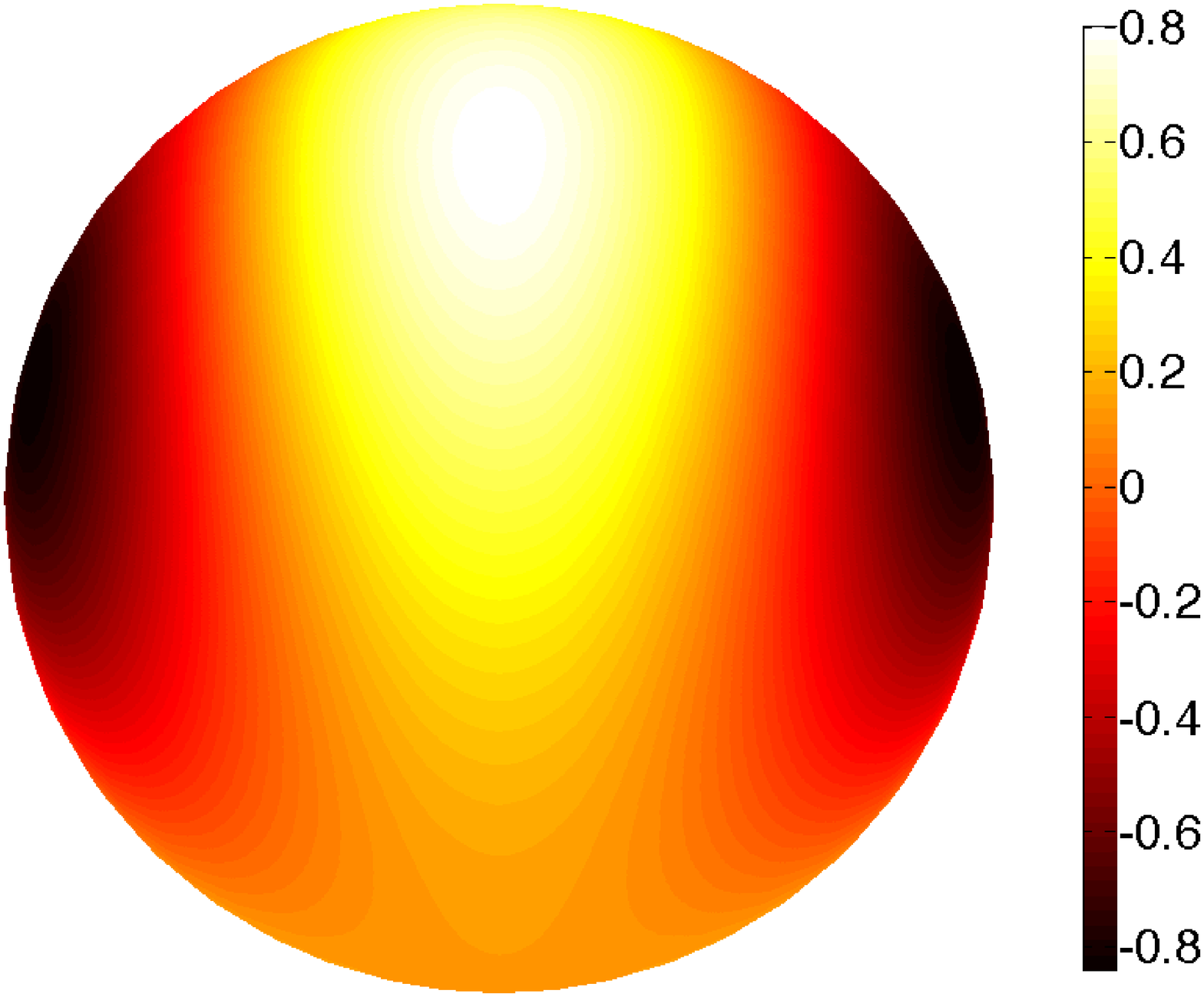}\includegraphics[width=4.2cm]{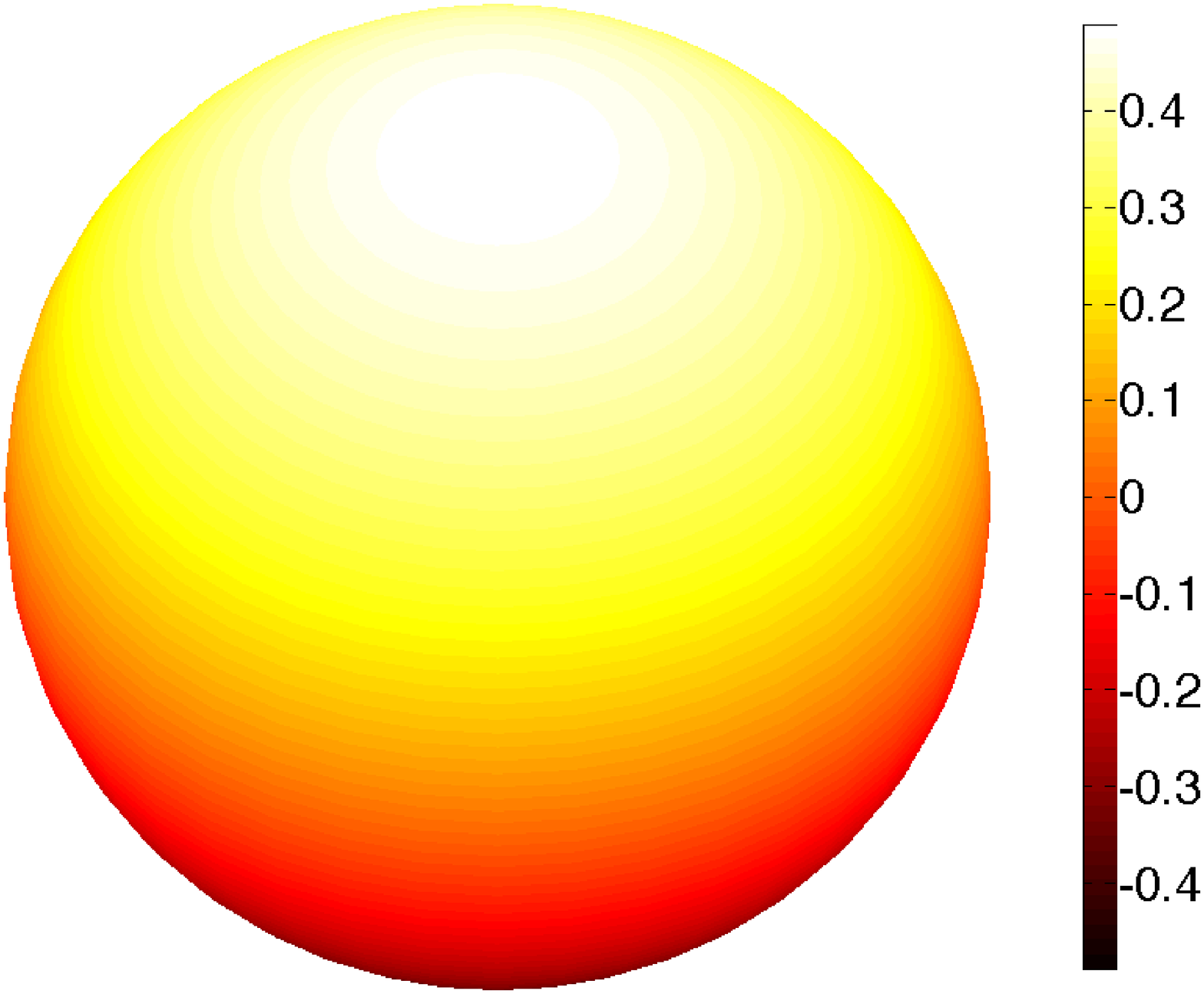}\end{center}

\caption{\label{fig:waveletplot}Plots of the real scale discretized wavelets
defined through relations (\ref{eq:sdw17}), (\ref{eq:sdw20}), and
(\ref{eq:sdw21}), at various analysis depths. A global band limit
$B=1024$ and a basis dilation factor $\alpha=2$ are chosen, as well
as an azimuthal band limit $N=3$ for the steerability. Light and
dark regions respectively correspond to positive and negative values
of the functions (see value bars). The wavelets are neither translated,
\emph{i.e.} they have their central position at the North pole, nor
rotated, \emph{i.e. }they are in their original orientation $\chi=0$
(the meridian $\varphi=0$ corresponds to a vertical line passing
by the North pole). The wavelets are represented at the four largest
analysis depths, $7\leq j\leq10=J_{B}(\alpha)$, identifying the four
largest scales. At $j=7$ (extreme left panel), $j=8$ (center-left
panel), and $j=9$ (center-right panel), the compact supports of the
scale discretized kernels respectively contain the frequencies $l=5$
to $l=15$ with a kernel maximum at $l=8$, $l=3$ to $l=7$ with
a kernel maximum at $l=4$, and $l=2$ to $l=3$ with a kernel maximum
at $l=2$. At $j=10$ (extreme right panel), the scale discretized
kernel only contains the frequency $l=1$. In real space, the dispersion
of angular distances around the central position on the sphere increases
with the analysis depth, in complete coherence with the constraint
(\ref{eq:kd13}). For the depths $j$ with $7\leq j\leq9$, the lowest
frequencies $l$ are greater or equal to $N-1=2$ and the azimuthal
frequency indices contained in the directional split are $m\in\{-2,0,2\}$.
These wavelets all have the same directionality property as measured
by an auto-correlation function evolving as $\cos^{2}(\Delta\chi)$.
For the depth $j=10$, the scale discretized wavelet is a pure dipole
($l=1$). The azimuthal frequency index is restricted to $m=0$, and
the wavelet is simply axisymmetric with a constant auto-correlation
function.}

\end{figure*}
As an illustration of the transition between the continuous and scale
discretized formalisms, we explicitly design a real scale discretized
factorized steerable wavelet $\Gamma$ with compact harmonic support,
from a real continuous wavelet $\Psi$. We firstly define the directional
split and kernel with generic values for the band limit $B$, for
the basis dilation factor $\alpha>1$, as well as for the azimuthal
band limit $N$ of steerability. We then illustrate the definition
for particular values.

The directionality coefficients of $\Psi$ and $\Gamma$ are identical
by the definition (\ref{eq:sdw2}). The steerability relation (\ref{eq:sdw3})
is imposed with an azimuthal band limit $N$. The function is imposed
to be real and to be even or odd both under rotation around itself
by $\pi$ and under a change of sign on $\varphi$. As discussed in
Subsection \ref{sub:Factorized-functions}, this corresponds to the
constraints that only the $T=N$ values $m\in T_{N}$ are allowed,
with $S_{lm}^{\Gamma*}=(-1)^{m}S_{l(-m)}^{\Gamma}$ , and $S_{lm}^{\Gamma}$
is real for even values of $N-1$, and purely imaginary for odd values
of $N-1$. One has $S_{00}^{\Gamma}=0$, and only the values $S_{lm}^{\Gamma}$
with $1\leq l<B$ and $0\leq m\leq l$ and $m\in T_{N}$ need to be
defined explicitly. These values are set in order to ensure a precise
structure of the auto-correlation function (\ref{eq:kd14}) under
the constraint (\ref{eq:sdw4}): \begin{equation}
S_{lm}^{\Gamma}=\eta_{N}\beta_{(N,m)}\left[\frac{1}{2^{\gamma_{(N,l)}}}\left({\gamma_{(N,l)}\atop \frac{\gamma_{(N,l)}-m}{2}}\right)\right]^{1/2},\label{eq:sdw17}\end{equation}
with $\eta_{N}=1$ for even values of $N-1$, $\eta_{N}=i$ for odd
values of $N-1$, $\beta_{(N,m)}=[1-(-1)^{N+m}]/2$, and $\gamma_{(N,l)}=\min(N-1,l-[1+(-1)^{N+l}]/2)$.
The auto-correlation function follows as \begin{equation}
C^{\Gamma}\left(\Delta\chi\right)=\sum_{l\in(\left\lfloor \alpha^{-1}B\right\rfloor ,B)}\tilde{K}_{\Gamma}^{2}(l)\cos^{\gamma_{(N,l)}}\left(\Delta\chi\right).\label{eq:sdw18}\end{equation}
When $N-1>\left\lfloor \alpha^{-1}B\right\rfloor +1$, the peakedness
of the auto-correlation is generically defined by the powers $\gamma_{(N,l)}$
of $\cos(\Delta\chi)$ at each value of $l$. This expresses the simple
fact that the azimuthal frequency index $m$ must always remain bounded
in absolute value by the overall frequency index: $|m|\leq l$. But
the values $\gamma_{(N,l)}$ ensure that the directionality coefficients
are independent of $l$ for $l\geq N-1$. Hence, when $N-1\leq\left\lfloor \alpha^{-1}B\right\rfloor +1$,
the auto-correlation function takes the form \begin{equation}
C^{\Gamma}\left(\Delta\chi\right)=\vert\vert\Gamma\vert\vert^{2}\cos^{(N-1)}\left(\Delta\chi\right),\label{eq:sdw19}\end{equation}
with $\vert\vert\Gamma\vert\vert^{2}=\sum_{l\in(\left\lfloor \alpha^{-1}B\right\rfloor ,B)}\tilde{K}_{\Gamma}^{2}(l)$.
Its peakedness increases as the power $N-1$ of $\cos(\Delta\chi)$.
The cost for the corresponding increase of directionality with $N$
is of course that a larger number of basis functions is required to
steer the wavelet.

Let us emphasize the importance of the structure (\ref{eq:sdw18})
in the global scheme of the scale discretized wavelet formalism. The
azimuthal band limit $N$ might be considered much smaller than the
lower bound of the compact harmonic support interval for the first
analysis depth $j=0$: $N\ll\left\lfloor \alpha^{-1}B\right\rfloor $.
However at each analysis depth $j\geq1$, the compact harmonic support
is defined in the interval $l\in(\left\lfloor \alpha^{-(1+j)}B\right\rfloor ,\left\lceil \alpha^{(1-j)}B\right\rceil )$.
Hence the structure (\ref{eq:sdw19}) of the auto-correlation function
breaks down to (\ref{eq:sdw18}) at a given analysis depth $j_{N}$,
defined as the lowest integer such that $N-1>\left\lfloor \alpha^{-(1+j_{N})}B\right\rfloor +1$.
If one wants to preserve the structure (\ref{eq:sdw19}) for all dilated
wavelets, the resolution of the identity (\ref{eq:sdw10}) can be
used up to a total analysis depth $J=j_{N}-1$.

The continuous kernel is defined from a Schwartz function with compact
support in the interval $(-1,1)$ on $\mathbb{R}$ as:\begin{eqnarray}
\tilde{K}_{\Psi}\left(k\right) & = & \exp\left[-\frac{1}{1-t^{2}\left(k\right)}\right]\quad\mbox{for}\quad t(k)\in(-1,1),\nonumber \\
\tilde{K}_{\Psi}\left(k\right) & = & 0\quad\mbox{for}\quad t(k)\notin(-1,1),\label{eq:sdw20}\end{eqnarray}
for the function \begin{equation}
t\left(k\right)=2\frac{\alpha k-B}{\left(\alpha-1\right)B}-1,\label{eq:sdw21}\end{equation}
which linearly maps the compact support interval $k\in(\alpha^{-1}B,B)$
onto $t\in(-1,1)$. The function $\tilde{K}_{\Psi}(k)$ is infinitely
differentiable for $k\in\mathbb{R}_{+}$. It notably exhibits a maximum
at the center $t(k)=0$ of the support interval and smoothly drops
down to zero at the interval bounds. Let us recall that the kernel
(\ref{eq:sdw20}) is by definition taken as a positive function. An
overall change of sign would simply flip the sign of the wavelet at
each point in real space. The scaling function $\tilde{\Phi}_{\Gamma}(k)$
and scale discretized kernel $\tilde{K}_{\Gamma}(k)$ follow from
relations (\ref{eq:sdw5}) and (\ref{eq:sdw7}) respectively. The
scaling function for values in the interval $k\in(\alpha^{-(1+j)}B,\alpha^{-j}B)$
for each analysis depth $j$ can be obtained by numerical integration.
Notice that the exactness of reconstruction provided by the formalism
is not affected by such a numerical integration, as long as the scale
discretized kernels are simply defined by differences of scaling functions
through relation (\ref{eq:sdw7}). Corresponding graphs are reported
in Figure \ref{fig:allkernels} for a band limit $B=1024$ and a basis
dilation factor $\alpha=2$ associated with a standard dyadic decomposition
of scales. 

Plots of the scale discretized wavelet are reported at various analysis
depths in Figure \ref{fig:waveletplot}, for $B=1024$, $\alpha=2$,
and for an azimuthal band limit $N=3$ for the steerability. These
plots notably illustrate localization and directionality properties
of the wavelet.

\subsection{Invertible filter bank}

\label{sub:Invertible-filter-bank}A scale discretized wavelet formalism
with relations (\ref{eq:sdw10}) and (\ref{eq:sdw12}) for factorized
steerable wavelets with compact harmonic support can be developed
by simply relying on a Littlewood-Paley decomposition, without any
contact with the continuous wavelet formalism. One simply needs to
choose any arbitrary scaling function satisfying relation (\ref{eq:sdw6})
and define the corresponding scale discretized kernels by differences
of scaling functions at successive scales.

Such invertible filter banks based on the harmonic dilation were already
developed in the case of axisymmetric wavelets \citep{starck06b},
and our definition of factorized steerable wavelets with compact harmonic
support allows a straightforward generalization to directional wavelets
with the kernel dilation. Also notice that the constraints of steerability
and compact harmonic support for the scale discretized wavelets can
technically be relaxed without affecting the Littlewood-Paley decomposition.
However both properties are essential for the control of localization
and directionality properties through kernel dilation. Moreover, in
the absence of compact harmonic support, the relation (\ref{eq:sdw10})
turns into a resolution of the contracted scaling function $\tilde{\Phi}_{\Gamma}^{2}(\alpha^{-1}l)$
which differs from unity below the band limit. In other words, the
filter bank developed in such a case analyzes the part of the signal
corresponding to its standard correlation with the contracted scaling
function, rather than the signal itself. In the absence of compact
harmonic support and steerability, essential multi-resolution properties
are also lost (see Subsection \ref{sub:Multi-resolution}). The memory
and computation time requirements of the algorithm for the analysis
and reconstruction of signals therefore increase significantly and
may rapidly become overwhelming.

Invertible filter banks based on the stereographic dilation have also
recently been proposed \citep{yeo06}, but they do not share these
essential multi-resolution properties.

\section{Exact multi-resolution algorithm}

\label{sec:Multiresolution-algorithm}In this section we identify
the multi-resolution properties of the scale discretized wavelet formalism
developed. We describe a corresponding algorithm for the analysis
and exact reconstruction of band-limited signals. We discuss in detail
the memory and computation time requirements of the algorithm. Finally,
an implementation of the algorithm is tested.

\subsection{Multi-resolution}

\label{sub:Multi-resolution}We consider the analysis and exact reconstruction
of a band-limited signal $F\in\textnormal{L}^{2}(\textnormal{S}^{2},\textnormal{d}\Omega)$
with a scale discretized wavelet $\Gamma\in\textnormal{L}^{2}(\textnormal{S}^{2},\textnormal{d}\Omega)$,
which is a factorized steerable function with compact harmonic support.
We consider a band limit $B$ and a basis dilation factor $\alpha>1$.

The signal is identified by $\mathcal{O}(B^{2})$ spherical harmonic
coefficients $\widehat{F}_{lm}$. Equivalently, sampled values $F(\omega_{i})$
of the signal on a number $\mathcal{O}(B^{2})$ of points $\omega_{i}$
are generally required in order to describe it completely. The integer
$i$ simply indexes the points of the chosen pixelization. Notably
exact quadrature rules for integration of band-limited signals on
$\textnormal{S}^{2}$ with band limit $B$ exist on equi-angular and
Gauss-Legendre pixelizations on $\mathcal{O}(B^{2})$ points. The
quadrature rules on HEALPix pixelizations on $\mathcal{O}(B^{2})$
points are non-exact but can be made very precise \citep{driscoll94,doroshkevich05a,gorski05}.

The compact harmonic support of the scale discretized wavelet $\Gamma_{\alpha^{j}}$
is reduced in the intervals $l\in(\left\lfloor \alpha^{-(1+j)}B\right\rfloor ,\left\lceil \alpha^{(1-j)}B\right\rceil )$
through the kernel dilation at each analysis depth $j$. As a function
on $\textnormal{SO(3)}$, the wavelet coefficients at depth $j$ exhibit
the same compact harmonic support as the scale discretized wavelet
$\Gamma_{\alpha^{j}}$. From relation (\ref{eq:sdw14}), the Wigner
$D$-transform $\widehat{(W_{\Gamma}^{F})}_{mn}^{l}(\alpha^{j})$
of the wavelet coefficients is indeed non-zero only in the same interval
as the wavelet. In particular, the band limit of the wavelet coefficients
is decreased to $\left\lceil \alpha^{(1-j)}B\right\rceil $ at depth
$j$. Consequently, the number of sampled values of the wavelet coefficients
is reduced at each increase of the analysis depths $j$ to $\alpha^{2(1-j)}\times\mathcal{O}(B^{2})$
discrete points of the form $(\omega_{0})_{i(j)}$ on $\textnormal{S}^{2}$,
where $i(j)$ simply indexes these points. The number of operations
required for their computation is reduced correspondingly. Hence,
the kernel dilation applied to scale discretized wavelets with compact
harmonic support provides a first strong multi-resolution property
for the formalism. 

The steerability of the wavelet is also important in the algorithmic
structure of the analysis \citep{wiaux05,wiaux06,wiaux07}, beyond
the fact that it ensures that directionality properties are preserved
through kernel dilation. Indeed, by linearity of the directional correlation
(\ref{eq:sdw13}), the general property of steerability (\ref{eq:kd6})
is transferred from the wavelet to the wavelet coefficients of any
signal. At each point $(\omega_{0})_{i(j)}$ and at each analysis
depth $j$, the wavelet coefficients of a signal $F$ with the scale
discretized wavelet $\Gamma_{\alpha^{j}}$ are known for all continuous
rotation angles $\chi\in[0,2\pi)$ as a linear combination of the
wavelet coefficients of $F$ with $M$ basis wavelets. As discussed,
the basis wavelets can be taken as specific rotations $\Gamma_{\chi_{p},\alpha^{j}}$
of the wavelet on itself by rotation angles $\chi_{p}\in[0,2\pi)$,
with interpolation weights given as simple translations by $\chi_{p}$
of a unique function $k(\chi)$. We consider wavelets for which the
number of rotations required can by optimized to $M=T\leq2N-1$, where
$T$ is the finite number of values of $m$ for which $\widehat{G}_{lm}$
has a non-zero value for at least one value of $l$. Consequently,
the steerability of the scale discretized wavelet $\Gamma_{\alpha^{j}}$
implies a reduction of the number of sampled values of the wavelet
coefficients to the $T$ values $\chi_{p}$ of the rotation angle,
with $0\leq p\leq T-1$, at each point $(\omega_{0})_{i(j)}$ and
at each analysis depth $j$. The number of operations required for
their computation is reduced correspondingly. From this perspective,
steerability provides a second strong multi-resolution property for
the formalism. All required sampled values of the wavelet coefficients
of a signal with a steerable wavelet may be mapped on a sphere for
each of the $T$ values of $\chi_{p}$, at each analysis depth $j$.

In summary, when multi-resolution properties of the formalism are
fully accounted for, a reduced number of discrete points of the form
$\rho_{I(j)}=((\omega_{0})_{i(j)},\chi_{p})$ on $\textnormal{SO(3)}$
are required for the sampled values $W_{\Gamma}^{F}(\rho_{I(j)},\alpha^{j})$
of the wavelet coefficients, where $I(j)=\{i(j),p\}$ simply indexes
these points at each analysis depth $j$.

\subsection{Algorithm}

\label{sub:Algorithm}The proposed algorithm works in harmonic space
on $\textnormal{S}^{2}$ and $\textnormal{SO(3)}$ in order to take
advantage of the directional correlation relation (\ref{eq:sdw14}).

Some precalculations are firstly required. The spherical harmonic
coefficients $\widehat{(\Gamma_{\alpha^{j}})}_{lm}$ of the scale
discretized wavelets must be designed at each analysis depth $j$.
A numerical integration can be required in order to compute the scaling
functions $\tilde{\Phi}_{\Gamma}^{2}(\alpha^{j}k)$ at all analysis
depths from the spherical harmonic coefficients $\widehat{\Psi}_{lm}$
of a continuous wavelet in relation (\ref{eq:sdw5}). The scale discretized
kernels $\tilde{K}_{\Gamma}^{2}(\alpha^{j}k)$ are then obtained by
differences of scaling functions, and multiplied by the directional
split chosen $S_{lm}^{\Gamma}$.

The analysis proceeds as follows. The band-limited signal $F$ is
given in terms of its sampled values $F(\omega_{i})$ on the $\mathcal{O}(B^{2})$
discrete points $\omega_{i}$ of $\textnormal{S}^{2}$. The spherical
harmonic coefficients $\widehat{F}_{lm}$ of the signal are computed
by quadrature through a direct spherical harmonic transform. The direct
Wigner $D$-function transform $\widehat{(W_{\Gamma}^{F})}_{mn}^{l}(\alpha^{j})$
of the wavelet coefficients is then simply obtained by the pointwise
product (\ref{eq:sdw14}). The computation of sampled values $W_{\Gamma}^{F}(\rho_{I(j)},\alpha^{j})$
of the wavelet coefficients requires an inverse Wigner $D$-function
transform at each analysis depth $j$. Before reconstruction, any
suitable analysis scheme can be applied on the wavelet coefficients,
for typical purposes of denoising or deconvolution. This provides
altered coefficients $\bar{W}_{\Gamma}^{F}(\rho_{I(j)},\alpha^{j})$.
The reconstruction proceeds trough the exact same operations as the
analysis, in reverse order. The Wigner $D$-function coefficients
$\widehat{(\bar{W}_{\Gamma}^{F})}_{mn}^{l}(\alpha^{j})$ of the altered
wavelet coefficients are computed by quadrature through a direct Wigner
$D$-function transform at each analysis depth $j$. The spherical
harmonic coefficients of the reconstructed signal $\widehat{\bar{F}}_{lm}$
are then obtained as a finite summation following from relations (\ref{eq:sdw15})
and (\ref{eq:sdw16}):

\begin{eqnarray}
\widehat{\bar{F}}_{lm} & = & \widehat{\left[\Phi_{\alpha^{J}}F\right]}_{lm}+\nonumber \\
 &  & \frac{2l+1}{8\pi^{2}}\sum_{j=0}^{J}\sum_{|n|\leq\min\left(N-1,l\right)}\widehat{\left(\Gamma_{\alpha^{j}}\right)}_{ln}\widehat{\left(\bar{W}_{\Gamma}^{F}\right)}_{mn}^{l}\left(\alpha^{j}\right),\nonumber \\
 &  & \,\label{eq:ma1}\end{eqnarray}
with\begin{equation}
\widehat{\left[\Phi_{\alpha^{J}}F\right]}_{lm}=\tilde{\Phi}_{\Gamma}^{2}\left(\alpha^{J}l\right)\widehat{F}_{lm}.\label{eq:ma2}\end{equation}
In the particular case where $J=J_{B}(\alpha)$, one gets trivially
$\widehat{\left[\Phi_{\alpha^{J}}F\right]}_{lm}=\delta_{l0}\delta_{m0}\widehat{F}_{00}$,
which corresponds to keep only the mean of the signal out of the analysis.

The samples $\bar{F}(\omega_{i})$ of the reconstructed signal are
recovered by simple inverse spherical harmonic transform. If no alteration
was applied to the wavelet coefficients, the exact same samples are
obtained as for the original signal $F$. This exactness also obviously
relies on the use of exact quadrature rules both for the direct spherical
harmonic transform of the signal in the analysis part, and for the
direct Wigner $D$-function transform of the wavelet coefficients
in the reconstruction part. This requires the choice of equi-angular
or Gauss-Legendre pixelizations on $\textnormal{S}^{2}$ defining
the discrete points $\omega_{i}$ for the sampling of the original
signal, and defining the discrete points $(\omega_{0})_{i(j)}$ for
the sampling the wavelet coefficients at each analysis depth $j$
and for each value $\chi_{p}$. Again HEALPix pixelizations provide
non-exact but very precise quadrature rules.

\subsection{Memory requirements}

\label{sub:Memory-requirements}We define the \emph{storage redundancy}
of the algorithm as the ratio of the number of sampled values of the
wavelet coefficients at all analysis depths with a scale discretized
wavelet, to the number of sampled values of the original signal itself.
A low storage redundancy is important for achieving as low memory
requirements as possible in a practical implementation of the algorithm.

The Wigner $D$-transform $\widehat{(W_{\Gamma}^{F})}_{mn}^{l}(\alpha^{j})$
of the wavelet coefficients is non-zero only in the interval $l\in(\left\lfloor \alpha^{-(1+j)}B\right\rfloor ,\left\lceil \alpha^{(1-j)}B\right\rceil )$
at each analysis depth $j$. Each frequency index $l$ is thus retained
exactly twice when all analysis depths $j$ are considered. Moreover,
for a steerable wavelet with azimuthal band limit $N$, the index
$n$ accounting for the wavelet directionality in relations (\ref{eq:sdw14})
and (\ref{eq:ma1}) takes by definition $T$ values, with $T\leq2N-1$.
On the contrary, the index $m$ is only related to the signal. Consequently,
the storage redundancy of the algorithm would be exactly $2T$ if
the wavelet coefficients were to be computed in harmonic space only.

However, the computation of the sampled values $W_{\Gamma}^{F}(\rho_{I(j)},\alpha^{j})$
of the wavelet coefficients in real space on the discrete points $\rho_{I(j)}=((\omega_{0})_{i(j)},\chi_{p})$
of $\textnormal{SO(3)}$ is of course essential for general analysis
purposes. Let us recall that a number $\alpha^{2(1-j)}\times\mathcal{O}(B^{2})$
of discrete points $(\omega_{0})_{i(j)}$ on $\textnormal{S}^{2}$
is required at each analysis depth $j$. This number of sampled values
is restricted by the band limit but not by the existence of a lower
bound of the compact harmonic support. Thanks to the steerability,
only $M=T$ values $\chi_{p}$ are required at each point $(\omega_{0})_{i(j)}$.
The storage redundancy of the algorithm is thus obtained by accounting
for the steerability and summing over all analysis depths $j$ with
$0\leq j\leq J$. In the most exacting case where $J=J_{B}(\alpha)$,
it simply reads as: \begin{equation}
\left[R_{\textnormal{s}}\right]_{(B,T)}\left(\alpha\right)=\left[1+c\left(\alpha^{2}\right)\left(1-\alpha^{-2J_{B}(\alpha)}\right)\right]T.\label{eq:ma3}\end{equation}
Let us recall that $c(\alpha^{2})=\alpha^{2}/(\alpha^{2}-1)\in[1,\infty)$
stands for the compactness of the scale discretized wavelet. The number
of the sampled values $W_{\Gamma}^{F}(\rho_{I(j)},\alpha^{j})$ of
the wavelet coefficients retained by the algorithm defines an order
of magnitude of the memory requirements, in units corresponding to
one coefficient per unit of memory, as:\begin{equation}
\left[M_{\textnormal{s}}\right]_{(B,T)}\left(\alpha\right)=\left[R_{\textnormal{s}}\right]_{(B,T)}\left(\alpha\right)\times\mathcal{O}(B^{2}).\label{eq:ma4}\end{equation}
For completeness, let us emphasize that this value accounts for the
memory requirements associated with the storage of the wavelet coefficients
only. Memory is also required for the storage of the $\mathcal{O}(B^{2})$
sampled values $F(\omega_{i})$ of the original signal and the $T\times\mathcal{O}(B)$
values of the spherical harmonic coefficients $\widehat{\Psi}_{lm}$
of the continuous wavelet, or equivalently of the spherical harmonic
coefficients $\widehat{(\Gamma_{\alpha^{j}})}_{lm}$ of the scale
discretized wavelets at all analysis depths $j$. Additional temporary
memory allocations are also necessary which depend on the precise
implementation of the algorithm. Hence, the value $[M_{\textnormal{s}}]_{(B,T)}\left(\alpha\right)$
is to be considered as a lower bound but still fixes an order of magnitude
for the memory requirements of the algorithm.

In the extreme case of a large basis dilation factor $\alpha\geq B$,
the compact harmonic support of the scale discretized wavelet essentially
gets as large as the band limit, with a compactness $c(\alpha^{2})\leq B^{2}/(B^{2}-1)$.
The maximum analysis depth goes to unity, $J_{B}(\alpha)=1$, which
implies that only two scales are required for the analysis. The storage
redundancy reaches its lowest value $[R_{\textnormal{s}}]_{(B,T)}(\alpha)=2T$.
As soon as $\alpha<B$, the harmonic support of the scale discretized
wavelet obviously gets more compact and more scales are required.
For values of the basis dilation factor very close to unity, the compactness
gets very high and may prevent the wavelets to be localized enough
in real space at the smallest analysis scale. A typically absurd value
$\alpha\leq[B/(B-1)]^{1/2}$ gives $c(\alpha^{2})\geq B$, which corresponds
to a compact harmonic support selecting at maximum one frequency at
a time. This simply reminds us of the fact that too high compactnesses
are prohibited in the framework of a wavelet analysis. Let us fix
ideas on practical intermediate values of $\alpha<B$. Notice that
the storage redundancy increases with the band limit $B$, as $J_{B}(\alpha)$
defined in (\ref{eq:sdw11}) obviously increases with $B$ for a fixed
value of $\alpha$. We give the upper bounds in the limit $B\rightarrow\infty$
and $J_{B}(\alpha)\rightarrow\infty$. A dyadic decomposition of the
scales\emph{ $\alpha=2$} corresponds to a compactness $c(4)=4/3$,
and the storage redundancy is bounded by $[R_{\textnormal{s}}]_{(B,T)}(2)\leq7T/3$
for any band limit $B$. A steerability relation with $T=3$ hence
gives a bound $[R_{\textnormal{s}}]_{(B,T)}(2)\leq7$. A more compact
support of the scale discretized wavelets set by $\alpha=1.1$ corresponds
to a compactness $c(1.21)\simeq6$, and the bound on the redundancy
rises to $[R_{\textnormal{s}}]_{(B,T)}(1.1)\lesssim7T$. A value $T=3$
then already gives $[R_{\textnormal{s}}]_{(B,T)}(1.1)\lesssim21$.

\subsection{Computation time requirements}

\label{sub:Computation-time-requirements}We define the \emph{computation
redundancy} of the algorithm as the ratio of the number of operations
required for the analysis and reconstruction of a signal at all analysis
depths with scale discretized wavelets, to the corresponding number
of operations at the first depth ($j=0$) and per azimuthal frequency
($T=1$, as for an axisymmetric wavelet which contains only $m=0$).
A low computation redundancy is essential for achieving as low computation
time requirements as possible.

The precalculation consists of the computation of the spherical harmonic
coefficients $\widehat{(\Gamma_{\alpha^{j}})}_{lm}$ of the scale
discretized wavelets from a continuous wavelet. At each analysis depth
$j$, the computation of the scale discretized kernel requires a one-dimensional
numerical integration of relation (\ref{eq:sdw5}). The corresponding
number of operations required is independent of $\alpha$ as each
real value $k\in[1,B)$ is covered exactly once by the continuous
wavelets at all analysis depths $j$, whose kernels have compact supports
in the intervals $k\in(\alpha^{-(1+j)}B,\alpha^{-j}B)$. The pointwise
product between the scale discretized kernel and the directional split
in (\ref{eq:sdw1}) requires $T\times\mathcal{O}(B)$ operations.
As it clearly appears in the following, the cost of these operations
is negligible relative to the cost of the analysis and reconstruction
themselves. Moreover, it must be performed only once for all signals
to be analyzed.

The analysis at a single analysis depth $j$ consists in a simple
directional correlation of $F$ with $\Gamma_{\alpha^{j}}$ on $\textnormal{S}^{2}$,
leading to the wavelet coefficients on $\textnormal{SO(3)}$. The
\emph{a priori} number of operations for a naive quadrature in relation
(\ref{eq:sdw13}) is of order $\alpha^{5(1-j)}\times\mathcal{O}(B^{5})$,
which becomes rapidly unaffordable. Fast directional correlation algorithms
based on relation (\ref{eq:sdw14}) and on the separation of the three
variables of integration on $\textnormal{SO(3)}$, were recently developed
\citep{wandelt01,wiaux06,mcewen07a,wiaux07}. They allow the exact
computation of the sampled values $W_{\Gamma}^{F}(\rho_{I(j)},\alpha^{j})$
of the wavelet coefficients at each analysis depth $j$ through a
number of operations at maximum of order $\alpha^{3(1-j)}T\times\mathcal{O}(B^{3})$.
This number of operations is mainly driven by the Wigner $D$-function
transform and naturally scales linearly with the number $M=T$ of
rotation angles $\chi_{p}$ required by the steerability of the wavelet.%
\footnote{For the spherical harmonic transform of the signal, fast algorithms
exist on equi-angular pixelizations \citep{driscoll94,healy03,healy04},
as well as on Gauss-Legendre \citep{doroshkevich05a,doroshkevich05b}
and HEALPix \citep{gorski05} pixelizations. The corresponding number
of operations required is at maximum of order $\alpha^{3(1-j)}\times\mathcal{O}(B^{3})$
thanks to the separation of the two variables of integration on $\textnormal{S}^{2}$.%
} The reconstruction is symmetric to the analysis and therefore requires
the same number of operations, in reverse order. The computation redundancy
of the algorithm is thus obtained by accounting for the steerability
and summing over all analysis depths $j$ with $0\leq j\leq J$. In
the most exacting case where $J=J_{B}(\alpha)$, it simply reads as:\begin{equation}
\left[R_{\textnormal{c}}\right]_{(B,T)}\left(\alpha\right)=\left[1+c\left(\alpha^{3}\right)\left(1-\alpha^{-3J_{B}(\alpha)}\right)\right]T.\label{eq:ma5}\end{equation}
 The number of operations required by the algorithm defines an order
of magnitude of the computation time requirements, in units corresponding
to one operation per unit of time, as: \begin{equation}
\left[T_{\textnormal{c}}\right]_{(B,T)}\left(\alpha\right)=\left[R_{\textnormal{c}}\right]_{(B,T)}\left(\alpha\right)\times\mathcal{O}(B^{3}).\label{eq:ma6}\end{equation}
In this expression, the impact of the compact harmonic support of
the scale discretized wavelet is concentrated in $c(\alpha^{3})=\alpha^{3}/(\alpha^{3}-1)\in[1,\infty)$.

Let us again fix ideas on practical intermediate values of $\alpha<B$,
and establish upper bounds in the limit $B\rightarrow\infty$ and
$J_{B}(\alpha)\rightarrow\infty$. A dyadic decomposition of the scales\emph{
$\alpha=2$} corresponds to a generalized compactness $c(8)=8/7$,
and the computation redundancy is bounded by $[R_{\textnormal{c}}]_{(B,T)}(2)=15T/7$
for any band limit $B$. A steerability relation with $T=3$ hence
gives a bound $[R_{\textnormal{c}}]_{(B,T)}(2)\leq45/7\simeq6.5$.
A more compact support of the scale discretized wavelets set by $\alpha=1.1$
corresponds to a generalized compactness $c(1.331)\simeq4$, and the
bound on the redundancy rises to $[R_{\textnormal{c}}]_{(B,T)}(1.1)\lesssim5T$.
A value $T=3$ then gives $[R_{\textnormal{c}}]_{(B,T)}(1.1)\lesssim15$.

\subsection{Implementation}

\label{sub:Implementation}%
\begin{table*}
\begin{center}\begin{tabular}{llllll}
\hline 
\noalign{\vskip\doublerulesep}
$B\qquad\qquad$ & $64\qquad\qquad$ & $128\qquad\qquad$ & $256\qquad\qquad$ & $512\qquad\qquad$ & $1024$\tabularnewline[\doublerulesep]
\hline
\noalign{\vskip\doublerulesep}
\noalign{\vskip\doublerulesep}
$\mu\,\,\textnormal{(MB)}\qquad\qquad$ & $1.3\qquad\qquad$ & $5.3\qquad\qquad$ & $21\qquad\qquad$ & $84\qquad\qquad$ & $340$\tabularnewline[\doublerulesep]
\noalign{\vskip\doublerulesep}
\noalign{\vskip\doublerulesep}
$\tau\,\,\textnormal{(min)}\qquad\qquad$  & $0.019\qquad\qquad$ & $0.092\qquad\qquad$ & $0.73\qquad\qquad$ & $7.0\qquad\qquad$ & $72$\tabularnewline[\doublerulesep]
\noalign{\vskip\doublerulesep}
\noalign{\vskip\doublerulesep}
$\epsilon\qquad\qquad$ & $8.6\times10^{-14}\qquad\qquad$ & $3.2\times10^{-13}\qquad\qquad$ & $8.9\times10^{-13}\qquad\qquad$ & $2.2\times10^{-12}\qquad\qquad$ & $7.4\times10^{-12}$\tabularnewline[\doublerulesep]
\hline
\noalign{\vskip\doublerulesep}
\end{tabular}\end{center}

\caption{\label{tab:implement}Test of the implementation of the proposed algorithm
for the analysis and reconstruction of signals on the sphere with
scale discretized wavelets. Memory used $\mu$ in Megabytes (MB),
as well as computation times $\tau$ in minutes (min) and numerical
errors $\epsilon$ both averaged over five random test signals are
reported, as measured on a $2.2$ GHz Intel Core 2 Duo CPU with $2$
Gigabytes of RAM. Five band limits are considered $B\in\{64,128,256,512,1024\}$,
and the basis dilation factor is set to $\alpha=2$. The signals are
decomposed up to the maximum analysis depths at each band limit. The
steerable wavelet used has an azimuthal band limit $N=3$, with only
the $T=3$ even values of the azimuthal index allowed: $m\in\{-2,0,2\}$.}

\end{table*}
The proposed algorithm was implemented and tested on a $2.2$ GHz
Intel Core 2 Duo CPU with $2$ Gigabytes of RAM. As already emphasized,
the choice of the pixelization on which the original signal $F$ is
sampled is essential to ensure the exactness or high precision of
the mere computation of its spherical harmonic coefficients and hence
of the whole analysis and reconstruction process. The exactness of
the proposed algorithm is simply tested by considering that the analysis
starts at the level of the spherical harmonic coefficients $\widehat{F}_{lm}$
of the original signal, and ends at the level of the spherical harmonic
coefficients $\widehat{\bar{F}}_{lm}$ of the reconstructed signal
$\bar{F}$. We also tested the memory and computation time requirements.
Let us recall that the corresponding contributions associated with
the removed direct spherical harmonic transform of the original signal
$F$ and inverse spherical harmonic transform leading to the reconstructed
signal $\bar{F}$ are overwhelmed by the inverse and direct Wigner
$D$-functions transforms required at each analysis depth $j$. 

Band limits $B\in\{64,128,256,512,1024\}$ are considered and the
basis dilation factor is set to $\alpha=2$, hence defining a typical
dyadic decomposition of scales. At each band limit, five test signals
are considered, directly defined through random spherical harmonic
coefficients $\widehat{F}_{lm}$ with independent real and imaginary
parts uniformly distributed in the interval $(-1,1)$. The steerable
wavelet $\Gamma$ defined and illustrated in Subsection \ref{sub:Example-wavelet-design}
is used, for an azimuthal band limit $N=3$. It only contains the
$T=3$ even values $m\in\{-2,0,2\}$. As discussed in Subsection \ref{sub:Factorized-functions},
the basis functions for the steerability can be chosen as the three
rotated versions $\Gamma_{\chi_{p}}$ with $\chi_{p}=\pi p/3$ for
$0\leq p\leq2$, and the function $k(\chi)$ follows accordingly.
The analysis is performed up to the maximum analysis depth for each
band limit: $J_{64}(2)=6$, $J_{128}(2)=7$, $J_{256}(2)=8$, $J_{512}(2)=9$,
and $J_{1024}(2)=10$. The numerical error associated with the algorithm
is evaluated as the maximum absolute value, for all values of $l$
and $m$, of the difference between the original and reconstructed
spherical harmonic coefficients: $\epsilon=\max_{l,m}|\widehat{F}_{lm}-\widehat{\bar{F}}_{lm}|$.
The algorithm is coded with double precision numbers, which sets the
unit of memory for the storage of coefficients to $8$ bytes.

The memory used $\mu$, as well as computation times $\tau$ and numerical
errors $\epsilon$ both averaged over the five random test signals
are reported in Table \ref{tab:implement}. The values reported respectively
illustrate the memory requirements (\ref{eq:ma4}), the computation
time requirements (\ref{eq:ma6}), and the exactness of reconstruction
of the proposed algorithm.

\section{Astrophysical application}

\label{sec:Astrophysical-application}In this section we emphasize
an important astrophysical application of the wavelet formalism defined
and implemented, discussing in some detail the issue of the detection
of cosmic strings through the denoising of full-sky CMB data. We firstly
introduce the question of the existence of topological defects in
the Universe. We highlight the non-Gaussianity of the component of
the CMB signal induced by cosmic strings and justify a wavelet decomposition
of the data as a way to enhance the sparsity of the wavelet coefficients
of the string signal. We then propose a denoising method based on
a statistical model of the wavelet coefficients of the string signal.
We also emphasize the need for a precise test allowing one to set
a confidence level on the string signal reconstructed from the denoised
wavelet coefficients.

\subsection{Topological defects}

Observations of the CMB and of the Large Scale Structure (LSS) of
the Universe have led to the definition of a concordance cosmological
model. The full-sky data of the WMAP experiment have played a dominant
role in developing this precise picture of the Universe \citep{bennett03,spergel03,hinshaw07,spergel07,hinshaw08,komatsu08}.
In this framework, the cosmic structures originate largely from Gaussian
adiabatic perturbations seeded in the early phase of inflation of
the Universe. However, cosmological scenarios motivated in the context
of theories unifying the fundamental interactions suggest the existence
of topological defects resulting from phase transitions at the end
of inflation. These defects would have participated to the formation
of the cosmic structures. While textures are more or less axisymmetric,
cosmic strings are a line-like version of defects \citep{vilenkin94,hindmarsh95,turok90}.
Even though observations largely fit with an origin of the cosmic
structures in terms adiabatic perturbations, room is still available
for the existence of a small fraction of topological defects. Moreover,
fundamental string theory predicts the existence of cosmic strings
in the {}``brane-world'' scenario \citep{davis05}. As a consequence,
the issue of the existence of topological defects represents today
a central question in cosmology.

Textures would induce hot and cold spots in the CMB with typical angular
sizes of several degrees on the celestial sphere \citep{turok90}.
A recent analysis \citep{cruz07} of the WMAP data showed that the
cold spot detected at $(\theta,\varphi)=(147^{\circ},209^{\circ})$
in Galactic spherical coordinates, is satisfactorily described by
a texture with an angular size around $10^{\circ}$. The main signature
of cosmic strings in the CMB is known as the Kaiser-Stebbins effect
\citep{kaiser84}, characterized by temperature steps along the strings,
with a typical angular size below $1^{\circ}$ on the celestial sphere.
Constraints have been set on a possible string contribution in terms
of upper limits on the so-called string tension $G\mu$, where $G$
stands for the gravitational constant. The string tension sets the
overall amplitude of the string contribution. These constraints mainly
come from the analysis of the string contribution to the overall CMB
angular power spectrum \citep{contaldi99,wyman05,wyman06,bevis08}.
Very few algorithms have been designed for the explicit identification
of cosmic strings through the Kaiser-Stebbins effect on full-sky data
\citep{jeong05,lo05}. No strong detection of cosmic strings has ever
been reported.

Current CMB experiments, among which WMAP, achieve an angular resolution
on the celestial sphere of the order of $10$ arcminutes, corresponding
to a limit frequency $B\simeq2\times10^{3}$. These experiments constrain
a possible string signal to be largely dominated by the standard Gaussian
CMB contribution at the frequencies $l$ probed, but it might nevertheless
become a dominant contribution at higher frequencies, due to the slow
decay of the corresponding angular power spectrum \citep{fraisse07,bevis08}.
The Planck experiment will provide full-sky CMB data at a resolution
of $5$ arcminutes, \emph{i.e.} with $B\simeq4\times10^{3}$ \citep{bouchet04}.
Important new information relative to a cosmic string signal will
therefore be available.

In this perspective, we sketch in the following a new statistical
approach \citep{wiaux08} for the identification and reconstruction
of cosmic strings through the denoising of full-sky CMB data. It is
specifically considered in the framework of the scale discretized
steerable wavelet formalism on the sphere%
\footnote{Notice that experiments such as the Arcminute Microkelvin Imager (AMI)
\citep{jones02,barker06}, the Atacama Cosmology Telescope (ACT) \citep{kosowsky06},
or the South Pole Telescope (SPT) \citep{ruhl04} will map the CMB
at a resolution around $1$ arcminute, \emph{i.e.} with $B\simeq2\times10^{4}$.
The corresponding prospects for the detection of strings are thus
improved relative to Planck data, but these experiments will provide
observations of small portions of the celestial sphere only. Specific
algorithms for the identification of cosmic strings in CMB data on
planar patches must be considered. As a formalism of scale discretized
steerable wavelets also exists on the plane \citep{simoncelli92},
it can also be used for the identification and reconstruction of cosmic
strings through the denoising of CMB data on planar patches, in a
statistical approach analogous to the one proposed below \citep{wiaux08}.%
}. Further refinement of this approach, as well as its precise implementation,
its application to CMB data, and its comparison with other detection
algorithms, are the subjects of a future work.

\subsection{Non-Gaussian string signal}

The standard component of the CMB signal induced by adiabatic perturbations
is a Gaussian signal on the sphere with a known angular power spectrum
in a given cosmological model. Topological defects, and in particular
cosmic strings, induce a non-Gaussian component of the CMB signal
with characteristic features defined at specific positions and scales.
The corresponding angular power spectrum exhibits a fixed characteristic
shape, with a slow decay at high frequencies. The complete non-Gaussian
statistical distribution of a string signal and the corresponding
angular power spectrum can indeed be deduced from simulations in the
chosen cosmological model \citep{fraisse07,bevis08}, up to an overall
amplitude of the string contribution as set by the unknown string
tension $G\mu$. These two statistically independent components simply
add linearly. We consider the perturbations of the signals around
their statistical mean. Instrumental Gaussian white noise with zero
mean also unavoidably adds as an independent component, setting the
limited sensitivity of the experiment considered. We leave apart any
issue of deconvolution of the experimental beam and also discard problems
of contamination of CMB data by foreground emissions. In the perspective
of the detection of cosmic strings, the non-Gaussian component from
strings represents the signal to be identified and reconstructed,
while the Gaussian components can be seen as a statistically independent
Gaussian noise. The overall signal $F$ reads as the sum of the string
signal and the noise in terms of a linear combination\begin{equation}
F\left(\omega_{i}\right)=a_{s}F_{s}\left(\omega_{i}\right)+F_{n}\left(\omega_{i}\right),\label{eq:aa1}\end{equation}
where $F_{s}$ represents the string signal for a string tension $a_{s}=G\mu$
normalized to unity, and $F_{n}$ represents the noise. The zero mean
signals $F$, $F_{s}$, and $F_{n}$ are considered to have a band
limit $B$, related to the resolution of the experiment under consideration,
and a number $\mathcal{O}(B^{2})$ of points $\omega_{i}$ are required
for their precise description.

In first approximation we can fix the cosmological parameters at their
values in the concordance cosmological model. This fixes the angular
power spectra of the noise $F_{n}$ and of the normalized string signal
$F_{s}$.

\subsection{Sparse wavelet coefficients}

Wavelets are by construction filters with zero mean (see (\ref{eq:sdw9})).
As such, they generically enhance discontinuities, and reduce smooth
patterns in the signal analyzed. The non-Gaussian string signal characterized
by temperature steps typically has a sparse expansion in terms of
wavelets. It indeed only exhibits a small number of wavelet coefficients
of large absolute value at the specific positions of the strings and
at their characteristic scales. On the contrary, the Gaussian contributions
are characterized by smooth patterns designed by their angular correlation
functions. Their expansion in terms of wavelets is not sparse. The
sparsity of the wavelet coefficients of the string signal justifies
the wavelet decomposition of the data. Moreover, directional wavelets\emph{
(i.e.} with an azimuthal band limit $N>1$) particularly apply for
an efficient detection of the localized directional features associated
with the Kaiser-Stebbins effect. Indeed, the more similar the filter
to the signal signatures, the better it magnifies these signatures.
Correspondingly, the detection of textures would more naturally follow
from an analysis with axisymmetric wavelets ($N=1$). Finally, as
emphasized already, the scale discretization of the wavelets is essential
for the reconstruction of the signal. In conclusion, the denoising
procedure will be more efficient when applied to the wavelet coefficients
of the signal observed $F$ decomposed with a scale discretized steerable
wavelet.

A suitable scale discretized steerable wavelet $\Gamma$ is thus chosen
with a given directionality set in terms of an azimuthal band limit
$N$. The basis dilation factor $\alpha$ and the total analysis depth
$J$ are also chosen, in order to optimize the number of analysis
depths in the range of frequencies $l$ concerned by the string signal.
Let us recall that the sampling of the wavelet coefficients of a signal
is defined on the points $\rho_{I(j)}=((\omega_{0})_{i(j)},\chi_{p})$
on $\textnormal{SO(3)}$. The value $I(j)=\{i(j),p\}$ simply indexes
these points at each analysis depth $j$ with $0\leq j\leq J\leq J_{B}(\alpha)$.
They correspond to the points $(\omega_{0})_{i(j)}$ on $\textnormal{S}^{2}$
for each value of the rotation angle $\chi_{p}$, with $0\leq p\leq T-1$,
as required by the steerability (see Subsection \ref{sub:Multi-resolution}).
By linearity of the wavelet decomposition (\ref{eq:sdw13}), the wavelet
coefficients $W_{\Gamma}^{F}$ of the overall signal $F$ read, in
terms of the wavelet coefficients of the normalized string signal
$F_{s}$ and of the noise $F_{n}$, as \begin{align}
W_{\Gamma}^{F}\left(\rho_{I(j)},\alpha^{j}\right) & =a_{s}W_{\Gamma}^{F_{s}}\left(\rho_{I(j)},\alpha^{j}\right)+W_{\Gamma}^{F_{n}}\left(\rho_{I(j)},\alpha^{j}\right),\label{eq:aa2}\end{align}
with $a_{s}W_{\Gamma}^{F_{s}}=W_{\Gamma}^{a_{s}F_{s}}$. The wavelet
coefficients $W_{\Gamma}^{F}$, $W_{\Gamma}^{F_{s}}$, and $W_{\Gamma}^{F_{n}}$
have zero statistical means just as the corresponding signals.

\subsection{Statistical model}

In a training phase, the statistical distributions of the wavelet
coefficients of a pure noise $F_{n}$ and of a pure normalized string
signal $F_{s}$ must be identified.

For the noise $F_{n}$, the wavelet coefficients remain Gaussian by
linearity. Assuming the statistical isotropy of the noise, the probability
density functions of the zero mean wavelet coefficients $W_{\Gamma}^{F_{n}}$
depend on the analysis depth $j$ but are independent of $\rho_{I(j)}$:\begin{equation}
f_{j}^{F_{n}}\left(W_{\Gamma}^{F_{n}}\right)\sim\exp\left[-\frac{1}{2}\left(\frac{W_{\Gamma}^{F_{n}}}{\sigma_{j}^{F_{n}}}\right)^{2}\right].\label{eq:aa3}\end{equation}
The variances $(\sigma_{j}^{F_{n}})^{2}$ can be inferred from the
known angular power spectrum of the noise in the range of frequencies
probed by the wavelets at the different analysis depths.

For the normalized string signal $F_{s}$, a Monte Carlo analysis
based on string signal simulations \citep{bevis08,fraisse07} is required
to fit a non-Gaussian model of the probability density function at
each depth $j$. The more reliable the simulations, the better the
model for the probability density functions. As a first approximation,
the computation of the variance and kurtosis of the wavelet coefficients
allows one to fit a generalized Gaussian distribution at each depth.
Assuming the statistical isotropy of the string signal, the probability
density functions of the zero mean wavelet coefficients $W_{\Gamma}^{F_{s}}$
again depend on the analysis depth $j$ but are independent of $\rho_{I(j)}$:\begin{equation}
f_{j}^{F_{s}}\left(W_{\Gamma}^{F_{s}}\right)\sim\exp\left[-\Big\vert\frac{W_{\Gamma}^{F_{s}}}{u_{j}}\Big\vert^{h_{j}}\right].\label{eq:aa4}\end{equation}
The parameters $u_{j}$ relate to the standard deviations $\sigma_{j}^{F_{s}}$
of the distributions. The corresponding variances $(\sigma_{j}^{F_{s}})^{2}$
reflect the angular power spectrum of the string signal in the range
of frequencies probed by the wavelets at the different analysis depths.
The parameters $h_{j}$ obviously measure the peakedness of the distributions,
and relate to their kurtoses. At the analysis depths corresponding
to the characteristic range of frequencies concerned by the string
signal, the sparsity of the wavelet coefficients can be associated
with peaked distributions $f_{j}^{F_{s}}$ with heavy tails (\emph{i.e.}
with kurtoses larger than $3$, or values $0<h_{j}<2$) relative to
a Gaussian distribution (\emph{i.e.} with a kurtosis equal to $3$,
or a value $h_{\textnormal{Gaussian}}=2$).

\subsection{Denoising}

The identification and reconstruction of a string signal from real
data can then be implemented as follows. 

Firstly, the overall signal $F$ is decomposed with the chosen scale
discretized steerable wavelet $\Gamma$, which gives the wavelet coefficients
$W_{\Gamma}^{F}$ through relation (\ref{eq:sdw13}).

Secondly, the string tension associated with a still hypothetical
string signal is estimated in relation (\ref{eq:aa1}). A precise
approach based on the analysis of the angular power spectrum of the
real data could be adopted, consisting in a likelihood analysis involving
all cosmological parameters including the string tension. This standard
approach was used to obtain the current constraints on the string
tension \citep{bevis08}, together with a reassessment of the other
cosmological parameters. The distributions $f_{j}^{F_{n}}$ and $f_{j}^{F_{s}}$
should then also be reassessed according to the modified angular power
spectra for $F_{n}$ and $F_{s}$, respectively. However, in the approximation
considered, all cosmological parameters are fixed at their values
in the concordance cosmological model throughout the analysis. The
angular power spectra of the noise $F_{n}$ and of the normalized
string signal $F_{s}$ are thus kept invariant. A rough estimation
of the string tension $a_{s}=G\mu$ is obtained from a least squares
method based on the variances of the wavelet coefficients of $F$,
$F_{s}$, and $F_{n}$ at all depths $j$. It is primarily intended
to serve to the denoising itself and not as a final estimation of
the string tension. The wavelet decomposition here simply helps to
bin the values of the power spectra before defining the constraints.
By statistical independence, the variances $(\sigma_{j}^{F})^{2}$
of the overall signal $F$ at each depth $j$ read, in terms of the
variances $(\sigma_{j}^{F_{s}})^{2}$ of the normalized string signal
$F_{s}$ and of the variances $(\sigma_{j}^{F_{n}})^{2}$ of the noise
$F_{n}$, as\begin{equation}
\left(\sigma_{j}^{F}\right)^{2}=a_{s}^{2}\left(\sigma_{j}^{F_{s}}\right)^{2}+\left(\sigma_{j}^{F_{n}}\right)^{2},\label{eq:aa5}\end{equation}
with $a_{s}^{2}(\sigma_{j}^{F_{s}})^{2}=(\sigma_{j}^{a_{s}F_{s}})^{2}$.
The wavelet coefficients $W_{\Gamma}^{F}$ are thus assumed to satisfy
relation (\ref{eq:aa2}) for an estimation $\bar{a}_{s}$ of the exact
value $a_{s}$. The statistical distributions $f_{j}^{F_{n}}$ for
the wavelet coefficients $W_{\Gamma}^{F_{n}}$ of the noise $F_{n}$
are identified as in (\ref{eq:aa3}). The statistical distributions
$f_{j}^{\bar{a}_{s}F_{s}}$ for the wavelet coefficients $W_{\Gamma}^{\bar{a}_{s}F_{s}}$
of the string signal $\bar{a}_{s}F_{s}$ are as in (\ref{eq:aa4})
with $u_{j}\rightarrow\bar{a}_{s}u_{j}$ and $h_{j}$ left invariant.
By Bayes' theorem, the posterior probability distribution function
$f_{j}^{(\bar{a}_{s}F_{s}\vert F)}$ at each depth $j$ for the wavelet
coefficients $W_{\Gamma}^{\bar{a}_{s}F_{s}}$ given the observed values
$W_{\Gamma}^{F}$ reads as\begin{eqnarray}
f_{j}^{(\bar{a}_{s}F_{s}\vert F)}\left(W_{\Gamma}^{\bar{a}_{s}F_{s}}\vert W_{\Gamma}^{F}\right) & \sim & f_{j}^{F_{n}}\left(W_{\Gamma}^{F}-W_{\Gamma}^{\bar{a}_{s}F_{s}}\right)\times\nonumber \\
 &  & f_{j}^{\bar{a}_{s}F_{s}}\left(W_{\Gamma}^{\bar{a}_{s}F_{s}}\right).\label{eq:aa6}\end{eqnarray}
Notice that one could also easily account for a flexibility in the
overall amplitude of the standard Gaussian component of the CMB, associated
with the cosmological parameter $\sigma_{8}$, in the same least squares
approach.

Thirdly, from the identified posterior probability, the wavelet coefficients
$W_{\Gamma}^{a_{s}F_{s}}$ of the string signal are estimated to values
$\bar{W}_{\Gamma}^{a_{s}F_{s}}$, separately at each point $\rho_{I(j)}$
for each analysis depth $j$. For example, in a maximum \emph{a posteriori}
approach, this estimation is defined as the value which maximizes
the posterior probability, while in a Bayesian least square approach,
it is defined as the expectation value of the posterior probability.

Finally, the estimated string signal $\overline{a_{s}F_{s}}$ is reconstructed
from the denoised wavelet coefficients $\bar{W}_{\Gamma}^{a_{s}F_{s}}$
through relations (\ref{eq:sdw15}) and (\ref{eq:sdw16}). The string
network imprinted in the analyzed CMB data is readily mapped as the
magnitude of gradient of the reconstructed signal.

\subsection{Confidence level of detection}

For the reasons discussed above, scale discretized steerable wavelets
should represent a very powerful tool for the identification and the
reconstruction of the string signal buried in the standard Gaussian
component of the CMB and in instrumental noise. The statistical approach
considered for the denoising procedure at each point $\rho_{I(j)}$
for each analysis depth $j$ specifically accounts for the shape of
the power spectra of the string signal and of the Gaussian components.
It also accounts for the peakedness of the non-Gaussian string signal,
characterized by a large kurtosis at each analysis depth $j$ characteristic
of the string signal.

After reconstruction, a hypothesis test can be set up in order to
assess if the estimated string network indeed arises from a string
signal with high probability. For example, the kurtosis of the magnitude
of gradient of the reconstructed signal can be compared to the corresponding
reconstructed kurtosis of combinations (\ref{eq:aa1}) of a string
signal with noise, through Monte Carlo analyses for various string
tensions. More statistics may also be combined, such as the variances
and kurtoses at different resolutions of the magnitude of gradient
of the reconstructed signal, in order to provide a more robust hypothesis
test. This procedure is also intended to provide an estimation of
the string tension $a_{s}=G\mu$ from the denoised signal, more precise
than the original estimation $\bar{a}_{s}$ which first served to
the denoising.

Let us finally notice that the Canny algorithm \citep{canny86} was
recently proposed for the detection of cosmic strings in CMB data
on planar patches \citep{amsel07}. It consists of an edge detection
in the map of the magnitude of gradient of the original signal, independently
of any denoising approach. This method could be implemented for the
analysis of full-sky CMB data, and compared to our algorithm in order
to assess their relative performances. In the context of our denoising
approach, this edge detection might actually be applied to the magnitude
of gradient of the denoised signal, in order to count the string segments
obtained, instead of computing the corresponding kurtosis. The confidence
level of the detection could then be assessed by comparison with the
corresponding counts for pure noise through a Monte Carlo analysis.

\section{Conclusion}

\label{sec:Conclusion}We have derived a scale discretized wavelet
formalism for the analysis and exact reconstruction of band-limited
signals on the sphere with directional wavelets. The combination of
the two properties of exact reconstruction and directionality was
lacking in the existing wavelet formalisms. As for the formalism developed
by \citet{antoine99} and \citet{wiaux05}, the translations of the
wavelets at any point on the sphere and their proper rotations are
still defined through the continuous three-dimensional rotations.
But the wavelets are factorized steerable functions with compact harmonic
support, and they are dilated through a kernel dilation directly defined
in harmonic space. 

As an intermediate step, a continuous wavelet formalism was obtained.
This by-product of our developments can be understood as an alternative
approach for the analysis of signals, with wavelets bearing new compact
harmonic support and directionality properties. However, the continuous
range of scales required for the analysis still prevents in practice
the exact reconstruction of the signals analyzed from their wavelet
coefficients.

The scale discretized wavelet formalism results from an integration
by slices of the dilation factor of the continuous formalism. It allows
in practice the exact reconstruction of band-limited signals from
their wavelet coefficients with a finite number of scales. It can
also be derived independently of the continuous wavelets, and can
be understood as a generalization of existing invertible filter bank
methods. The multi-resolution properties of the formalism were identified
and a corresponding exact algorithm was described. The memory and
computation time requirements were discussed and an implementation
was tested.

This formalism is of interest in a large variety of fields, notably
for the denoising or the deconvolution of signals on the sphere with
a sparse expansion in wavelets. It typically concerns signals identified
by directional features at specific positions and scales. In astrophysics,
it finds a particular application for the identification of localized
directional features in CMB data, such as the imprint of topological
defects, in particular cosmic strings, and for their reconstruction
after separation from the other signal components. In this context,
we have discussed a new statistical approach for the detection of
cosmic strings through the denoising of full-sky CMB data. This application
is the subject of a future work.

\section*{Acknowledgments}

The authors wish to thank L. Jacques for valuable comments. The work
of Y. W. is funded by the Swiss National Science Foundation (SNF)
under contract No. 200020-113353. Y. W. is also a Postdoctoral Researcher
of the Belgian National Science Foundation (F.R.S.-FNRS). J. D. M.
is a Research Fellow of Clare College, Cambridge.

\label{lastpage}

\end{document}